\documentclass[11pt,a4paper]{article}

\usepackage{graphicx}  

\usepackage{amsmath,amsfonts,amssymb}  
\usepackage{color}
\usepackage{url}

\newcommand{\be}{\begin{equation}}
\newcommand{\ee}{\end{equation}}
\newcommand{\ra}{\Rightarrow}
\newcommand{\csch}{\mbox{csch}}
\newtheorem{prop}{Proposition}

\begin{document}

\begin{center}
{\huge Periodic solutions for a pair of delay-coupled \\ excitable theta neurons}

\vspace*{5mm}

{\large Carlo R. Laing} \\[1mm]

School of Mathematical and Computational Sciences, 
Massey University, \\ Private Bag 102-904, North Shore Mail Centre, 
Auckland 0745, New Zealand

\vspace*{5mm}

{\large Bernd Krauskopf} \\[1mm]

Department of Mathematics, The University of Auckland, \\ Private Bag 92019, Auckland 1142, New Zealand \\[5mm]

\end{center}

\vspace*{-2mm}

\begin{center}
{\large November 2024}
\end{center}

\vspace*{-3mm}

\begin{abstract}
We consider a pair of identical theta neurons in the excitable regime, each coupled to the other via a delayed Dirac delta function with the same delay. This simple network can support different periodic solutions, and we concentrate on two important types: those for which the neurons are perfectly synchronous, and those where the neurons are exactly half a period out of phase and fire alternatingly. Owing to the specific type of pulsatile feedback, we are able to determine these solutions and their stability analytically. More specifically, (infinitely many) branches of periodic solutions of either type are created at saddle-node bifurcations, and they gain stability at symmetry-breaking bifurcations when their period as a function of delay is at its minimum. We also determine the respective branches of symmetry-broken periodic solutions and show that they are all unstable. We demonstrate by considering smoothed pulse-like coupling that the special case of the Dirac delta function can be seen as a sort of normal form: the basic structure of the different periodic solutions of the two theta neurons is preserved, but there may be additional changes of stability along the different branches. 
\end{abstract}

\section{Introduction}

Excitability is a common phenomenon in nonlinear 
dynamical systems~\cite{IzhikevichBook,exciteoverview}. An excitable system typically
has a stable resting state and, when subject to a small perturbation, stays near and relaxes to this resting state. However, for a sufficiently large perturbation above what is known as the \emph{excitability threshold}, the system has a stereotypical response (such as a neuron firing an action potential or a laser emitting a pulse of light) before returning to the rest state; here the magnitude of the response is largely independent of the size of the perturbation, once it is above the excitability threshold. 

We consider here the prototypical example known as the \emph{theta neuron}, which is a mathematical model for the single angular variable $\theta(t) \in (-\pi,\pi]$, given by 
\be
   \frac{d\theta}{dt}=1-\cos{\theta}+(1+\cos{\theta})I. \label{eq:thetaneuron}
\ee
The theta neuron is the normal form of the saddle-node-on-invariant-circle (SNIC) bifurcation~\cite{ermkop86,erm96} and has the advantage that its dynamics can be found explicitly for constant $I$. In particular, this system is excitable when $I<0$, in which case it has two equilibria at $\theta_{\pm}=\pm 2\tan^{-1}{(\sqrt{-I})}$, where $\theta_-$ is the stable rest state and $\theta_+$ is unstable and forms the excitability threshold. A perturbation applied at $\theta_-$ that moves $\theta$ above $\theta_+$ results in $\theta$ moving through $\pi$ before returning to $\theta_-$, and (by convention) this is the moment when the theta neuron `fires' and produces a pulse. Note that, under the transformation $V=\tan{(\theta/2)}$, the theta neuron is equivalent to the quadratic integrate-and-fire neuron with the rule $V(t^+)=-\infty$ if $V(t^-)=\infty$, that is, $V$ is reset when the neuron fires.

An excitable unit, such as a neuron cell or excitable optical element, is able to react to inputs with the generation of a pulse, depending on the strength and timing of the inputs. Whenever a number of neurons or other elements are coupled to themselves and/or to each other, there are necessarily delays in when they receive input and how fast they react to it~\cite{pophau06}. Such delays may play an important role and, more generally, they are ubiquitous in dynamical systems. In particular, it is of interest to study the behavior of excitable systems with delayed interactions, where delays range from synaptic processing delays in neural systems~\cite{lailon03} to delays between coupled optical systems, which arise even when signals travel at the speed of light~\cite{KelleherPRE10}. 

In our previous work~\cite{laikra22} we studied a single theta neuron with delayed self-coupling given by a Dirac delta function of time. This specific type of pulsatile feedback allows us to analytically determine the dynamics between times at which the feedback acts, which can be described equivalently by a discrete map for the spike times. This allowed us to find explicit expressions for all periodic solutions, their stability and bifurcations of this excitable system with self-feedback. Our results show that the theta neuron with delayed delta-function self-coupling can be considered as a sort of normal form: while being fully treatable analytically, it captures the essentials of observed self-sustained oscillations in excitable systems with delayed self-feedback, including laser systems \cite{GarbinNC15, KelleherPRE10, KrauskopfWalker, RomeiraNSR16, TerrienPRA17, Terrien2018OL} and an actual cell~\cite{wedslo21}.

In this paper, we consider two identical theta neurons that are delay-coupled
via Dirac delta functions to one another. To this end, we set $I=-1$ from now on without loss of generality and study the system 
\begin{align}
   \frac{d\theta_1}{dt} & =1-\cos{\theta_1}+(1+\cos{\theta_1})\left(-1+\kappa\sum_{i:t-\tau<s_i<t} \delta(t-s_i-\tau)\right), \label{eq:dth1} \\
    \frac{d\theta_2}{dt} & =1-\cos{\theta_2}+(1+\cos{\theta_2})\left(-1+\kappa\sum_{i:t-\tau<t_i<t} \delta(t-t_i-\tau)\right). \label{eq:dth2}
\end{align}
Here, $\tau$ is the (constant) delay and $\kappa$ is the strength of coupling between the neurons, the firing times in the past of neuron 1 are $\{\dots, t_{-3},t_{-2},t_{-1},t_0\}$ and those of neuron 2 are $\{\dots, s_{-3},s_{-2},s_{-1},s_0\}$. Moreover, the influence of the delta function is to increment $\theta$ according to
\be
   \tan{(\theta^+/2)}=\tan{(\theta^-/2)}+\kappa, \label{eq:reset}
\ee
where $\theta^-$ and $\theta^+$ are the values of $\theta$ before and after the delta function acts, respectively~\cite{laikra22}. 

System~\eqref{eq:dth1}--\eqref{eq:dth2} constitutes the next step up in complexity from the case of a delay-self-coupled theta neuron studied in~\cite{laikra22} and, at the same time, it is a normal form for the simplest case of a network of only two delay-coupled excitable cells or elements. Pairs of oscillators of different type with delayed coupling have been studied commonly~\cite{schwag89,dodsen04,yeustr99}. The case of two delay-coupled excitable cells or elements, however, appears to have received less attention. We now discuss previous work on oscillations in such a system, which mostly concerns delay-coupled FitzHugh-Nagumo (FHN) systems. In~\cite{dahhil09,schhil09} two FHN systems with diffusive delay-coupling are considered. A stable periodic solution is found, where the individual FHN systems oscillate out of phase with one another, with a period slightly greater than twice the delay; this solution disappears in a saddle-node bifurcation as the delay is decreased. A pair of delay-coupled FHN systems is studied in~\cite{sonxu12}, and in-phase and anti-phase periodic solutions are found, albeit for a parameter regime where the uncoupled FHN systems have periodic solutions, rather than being excitable. Both in-phase and anti-phase periodic solutions, with periods roughly equal to the delay and twice the delay, respectively, are also found in~\cite{weiern14} for a pair of modified FHN systems. Two delay-coupled excitable FHN systems are studied in~\cite{burtod03}, where both synchronous and alternating period solutions are found  and the existence of less-symmetric periodic solutions is also mentioned. A living coupled oscillator system is constructed in~\cite{takfuj00} by using a plasmodial slime mould; depending on (effectively) the delay and the coupling strength between the two sub-parts, both synchronous and alternating periodic solutions are found. 

We show here that our approach and techniques can be used to find all periodic solutions of system~\eqref{eq:dth1}--\eqref{eq:dth2}, which may have differing spatio-temporal symmetries in light of the symmetry of exchanging the two theta neurons. Explicit expressions for $\theta_{1,2}$ between the times at which the delta function acts, in conjunction with the effect of the delta-function coupling~\eqref{eq:reset}, allow us to construct these periodic solutions. More specifically, we derive a map for the next firing times and linearise it about a periodic solution to obtain analytic expressions for its stability. This technique is similar to that in~\cite{klinek11}, with the main difference being that they consider oscillators for which $d\theta/dt$ is constant between the times at which the coupling acts, so that when uncoupled an oscillator will fire periodically. In contrast, the excitable neurons we consider will fire at most once when uncoupled. We also mention in this context the work of~\cite{lilin17}, who consider a network of three phase oscillators coupled by delayed Dirac delta functions.

The paper is structured as follows. We analyse synchronous (in-phase) periodic solutions in Sec.~\ref{sec:synch}, where we determine conditions for their existence and stability, and also describe symmetry-broken solutions that branch from them. In Sec.~\ref{sec:alt}, we perform a similar analysis for alternating (anti-phase) periodic solutions for which the neurons fire half a period out of phase with one another. A numerical bifurcation study in Sec.~\ref{sec:smooth} shows that the basic structure of the different periodic solutions is preserved when the delayed coupling is smooth rather than given by a delta function; however, there may be additional changes of stability along branches of periodic solutions. We conclude in Sec. ~\ref{sec:conclusions}, where we also briefly mention directions for future research. The appendix presents the arguments to establish the instability of the symmetry-broken solutions that bifurcate from the synchronous and the alternating periodic solutions.

\section{Synchronous solutions}
\label{sec:synch}

Since we set $I = -1$, each individual theta neuron, as described by~\eqref{eq:thetaneuron} with variable $\theta$ (standing for $\theta_1$ and/or $\theta_2$) has its equilibria at $\theta_\pm = \pm \frac{\pi}{2}$. If an initial condition $\theta(0)$ satisfies $\theta_+ = \frac{\pi}{2} < \theta(0)$ then the solution of~\eqref{eq:thetaneuron} (with $I = -1$) is
\be \theta(t)=2\tan^{-1}{\left[-\coth{\left(t-\coth^{-1}{\left[\tan{\left(\frac{\theta(0)}{2}\right)}\right]}\right)}\right]}. \label{eq:solA}
\ee
For our analysis in this and the next section we need two special cases of this explicit solution. 
If the neuron has fired at $t=0$, i.e.~$\theta(0)=\pi$, then
\be
   \theta(t)=2\tan^{-1}{\left[-\coth{(t)}\right]}, \label{eq:solB}
\ee
where $t$ is the time since firing. Moreover, when $\theta(0)>\pi/2$, the time $\Delta$ until firing satisfies
\[ \pi=2\tan^{-1}{\left[-\coth{\left(\Delta-\coth^{-1}{\left[\tan{\left(\frac{\theta(0)}{2}\right)}\right]}\right)}\right]}, 
\]
which reduces to
\be
   \Delta=\coth^{-1}{\left[\tan{\left(\frac{\theta(0)}{2}\right)}\right]}. \label{eq:Delta}
\ee

An important property of system~\eqref{eq:dth1}--\eqref{eq:dth2} is its reflection symmetry of interchanging the two theta neurons, given formally by $(\theta_1,\theta_2) \mapsto (\theta_2,\theta_1)$. General theory~\cite{golshaefer} implies the possibility of periodic solutions with different spatio-temporal symmetries, including those where the two neuron are in phase, in anti-phase, or have an intermediate-phase between them. Note also that the solution $\theta_1=\theta_2=3\pi/2$ is a stable fixed point 
of~\eqref{eq:dth1}--\eqref{eq:dth2} for any $\kappa$. Thus, to observe periodic solutions there must have been some firing of at least one of the neurons in the past (for example, due to an external stimulus).

In this section, we consider the first type of periodic solution of system~\eqref{eq:dth1}--\eqref{eq:dth2}, where the neurons are in phase, that is, they are perfectly synchronous.  The influence of one neuron on the other is then the same as that of the neuron on itself. While the existence of such synchronous solutions is governed by the equations of a single neuron with delayed self-coupled~\cite{laikra22}, their stability is more complicated. Namely, there is now the possibility of bifurcations out of the symmetry subspace (where the two phases are equal), which creates solutions for which the neurons have a phase difference and are, hence, no longer perfectly synchronised. The results in this section are summarized by Fig.~\ref{fig:synchper}, which shows the branches of synchronous periodic solutions, their stability and bifurcations, as well as branches of bifurcating symmetry-broken periodic solutions.

\subsection{Existence and stability of synchronous periodic solutions}
\label{sec:stab_inphase}

As shown in~\cite{laikra22}, perfectly synchronous periodic solutions of system~\eqref{eq:dth1}--\eqref{eq:dth2} with period $T$ satisfy
\be
   \coth{[(n+1)T-\tau]}=\kappa+\coth{[nT-\tau]}, \label{eq:existSA}
\ee
where $n$ is the number of past firing times in the interval $(-\tau,0)$, assuming that
a neuron has just fired at time $t=0$. The primary branch, corresponding to $n=0$,
is given explicitly by
\be
   T(\tau)=\tau+\coth^{-1}{(\kappa-\coth{\tau})}. \label{eq:existS}
\ee
Secondary branches are given parametrically, by using the reappearance of periodic
solutions in delay differential equations with fixed delays~\cite{yanper09}, as
\be
  (\tau,T)=(s+nT(s),T(s)), \label{eq:para}
\ee
where $\coth^{-1}{(\kappa-1)}<s<\infty$. 

\begin{figure}[t!]
\begin{center}
\includegraphics[height=10.1cm]{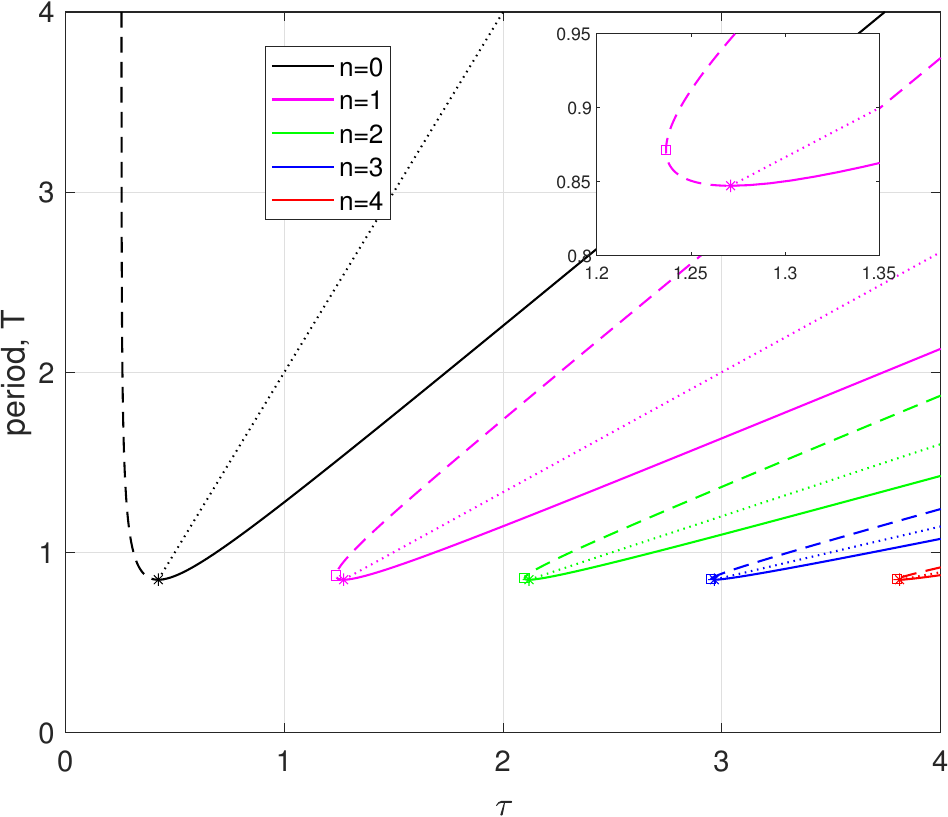}
\caption{Branches of synchronous and corresponding symmetry-broken periodic solutions of system~\eqref{eq:dth1}--\eqref{eq:dth2} with $n=0,1,2,3,4$. Solid curves are stable (to the right of each minimum) and dashed curves are unstable. Saddle-node bifurcations are marked with squares, symmetry-breaking bifurcations with stars, and bifurcating pairs of symmetry-broken solutions are represented by dotted lines. The inset is an enlargement of the branch with $n=1$. Here, $\kappa=5$.}
\label{fig:synchper}
\end{center}
\end{figure}

These branches of synchronous periodic solutions are shown in Fig.~\ref{fig:synchper} for $n=0,1,2,3,4$ and $\kappa=5$: they are the tilted parabola shaped curves that are also shown in~\cite[Fig.~3]{laikra22} for the case of a single theta neuron. However, the stability of the periodic solutions is different for the case of two synchronous theta neurons, and this is seen as follows.

Suppose neuron 1 last fired at time $t_0$ and neuron 2 last fired at $s_0$ where $s_0\approx t_0$. The most distant past firing of neuron 1 in $(t_0-\tau,t_0)$ is $t_{-n}$ and the most distant past firing of neuron 2 in $(s_0-\tau,s_0)$ is $s_{-n}$. After $t_0$, neuron 1 has its phase incremented for the next time at $\tau-(t_0-s_{-n})$ due to a past firing of neuron 2. Before the reset, from~\eqref{eq:solB}, $\theta_1$ equals
\[
   \theta_1^-=2\tan^{-1}{[-\coth{(\tau-(t_0-s_{-n}))}]}
\]
and after reset it is $\theta_1^+$, where
\[
   \tan{(\theta_1^+/2)}=\tan{(\theta_1^-/2)}+\kappa. 
\]
Neuron 1 will then fire after a further time $\Delta_1$ where, from~\eqref{eq:Delta},
\[
   \Delta_1=\coth^{-1}{[\tan{(\theta_1^+/2)}]}.
\]
Thus, we conclude that $t_1=t_0+\tau-(t_0-s_{-n})+\Delta_1$.

The argument for neuron 2 is the same: its phase is incremented for the first time at $\tau-(s_0-t_{-n})$, and the analogous expressions for $\theta_2^-$ and $\theta_2^+$ mean that neuron 2 will then fire after the further time 
\[
   \Delta_2=\coth^{-1}{[\tan{(\theta_2^+/2)}]},
\]
yielding $s_1=s_0+\tau-(s_0-t_{-n})+\Delta_2$.

Hence, we have
\begin{align*}
   s_1 & =\tau+t_{-n}+\coth^{-1}{[\tan{(\theta_2^-/2)}+\kappa]}=\tau+t_{-n}+\coth^{-1}{[\kappa-\coth{(\tau-s_0+t_{-n})}]}, \\
   t_1 & =\tau+s_{-n}+\coth^{-1}{[\tan{(\theta_1^-/2)}+\kappa]}=\tau+s_{-n}+\coth^{-1}{[\kappa-\coth{(\tau-t_0+s_{-n})}]}, 
\end{align*}
which give $s_1$ and $t_1$ in terms of previous firing times. By considering this argument for different $i$, we obtain the general case
\begin{align}
   s_{i+1} & =\tau+t_{i-n}+\coth^{-1}{[\kappa-\coth{(\tau-s_i+t_{i-n})}]}, \label{eq:spsynch} \\
   t_{i+1} & =\tau+s_{i-n}+\coth^{-1}{[\kappa-\coth{(\tau-t_i+s_{i-n})}]},  \label{eq:tpsynch}
\end{align}
which we write in the form 
\begin{align*}
   F(s_{i+1},t_{i-n},s_i) & = 0, \\
   G(t_{i+1},s_{i-n},t_i) & = 0.
\end{align*}
To determine stability we perturb $t_i\to t_i+\eta_i$ and $s_i\to s_i+\mu_i$, which gives to linear order
\begin{align}
  \frac{\partial F}{\partial s_{i+1}}\mu_{i+1}+\frac{\partial F}{\partial t_{i-n}}\eta_{i-n}+\frac{\partial F}{\partial s_{i}}\mu_{i} & =0, \label{eq:Fpartial}\\
  \frac{\partial G}{\partial t_{i+1}}\eta_{i+1}+\frac{\partial G}{\partial s_{i-n}}\mu_{i-n}+\frac{\partial G}{\partial t_{i}}\eta_{i} & =0. \label{eq:Gpartial}
\end{align}
When evaluating the partial derivatives at a periodic solution with period $T$, this becomes
\begin{align}
  -\mu_{i+1}+(1-\gamma)\eta_{i-n}+\gamma\mu_{i}=0, \label{eq:mu} \\
  -\eta_{i+1}+(1-\gamma)\mu_{i-n}+\gamma\eta_{i}=0,  \label{eq:eta}
\end{align}
where
\be
   \gamma=\frac{\coth^2{(\tau-nT)}-1}{[\kappa-\coth{(\tau-nT)}]^2-1}. \label{eq:gammaA}
\ee
From arguments in~\cite{laikra22} we know that $\gamma >0$.

Since~\eqref{eq:mu}--\eqref{eq:eta} are constant coefficient linear difference equations, 
we assume that solutions have the form $\eta_i=A\lambda^i$ and $\mu_i=B\lambda^i$ for some constants $A$ and $B$, giving 
\be
  \begin{pmatrix} (1-\gamma)\lambda^{i-n} & \gamma\lambda^i-\lambda^{i+1} \\
  \gamma\lambda^i-\lambda^{i+1} & (1-\gamma)\lambda^{i-n} \end{pmatrix} \begin{pmatrix} A \\ B \end{pmatrix}=\begin{pmatrix} 0 \\ 0 \end{pmatrix} \label{eq:matrix}
\ee
when written in matrix form. To have nontrivial solutions of~\eqref{eq:matrix} 
we need the determinant 
\[ (1-\gamma)^2\lambda^{2i-2n}-\gamma^2\lambda^{2i}+2\gamma\lambda^{2i+1}-\lambda^{2i+2} 
\]
of the matrix in~\eqref{eq:matrix} to be zero. Multiplying by $-\lambda^{2n-2i}$ gives the characteristic equation
\be
   D_s(\lambda) :=  \lambda^{2n}(\lambda-\gamma)^2-(1-\gamma)^2=0 \label{eq:charA}
\ee
of the synchronous periodic solutions. The roots of $D_s(\lambda)$ are the Floquet multipliers of the synchronous periodic solution with given $n$, and they depend on the parameter $\gamma > 0$ from \eqref{eq:gammaA}. Hence, the periodic solution is stable when all roots of~$D_s(\lambda)$ are smaller than 1 in modulus, and a bifurcation occurs when there is a root $\lambda$ with $|\lambda| = 1$. Note that~$D_s(\lambda)$ in~\eqref{eq:charA} factors as $D_s(\lambda)=g_s(\lambda)\tilde{g}_s(\lambda)=0$ with 
\begin{align}
   g_s(\lambda) & =\lambda^n(\lambda-\gamma)-(1-\gamma), \label{eq:gg}\\
   \tilde{g}_s(\lambda) & = \lambda^n(\lambda-\gamma)+(1-\gamma). \label{eq:ggtilde}
\end{align}
This is a reflection of the symmetry of system~\eqref{eq:dth1}--\eqref{eq:dth2}: $g_s(\lambda)=0$ is actually the characteristic equation for a single self-coupled neuron~\cite{laikra22}, which gives the stability inside the symmetry subspace with $\theta_1 = \theta_2$; and the factor $\tilde{g}_s(\lambda)$ accounts for the stability in the direction transverse to the symmetry subspace. 

The properties of $g_s$ are known from~\cite[Proposition~1]{laikra22}. For $0<\gamma<(n+1)/n$ all roots of $g_s$ have magnitude less than 1, and a single root leaves the unit circle with positive speed at $\lambda = 1$ as $\gamma$ increases through $(n+1)/n$; this is a saddle-node bifurcation of the synchronous periodic solution. The overall stability of the synchronous periodic solution also concerns the roots of $\tilde{g}_s$, for which we have the following result.

\begin{prop}[Properties of the roots of $\tilde{g}_s$]
\verb+ + \\[-8mm]  
\label{prop:g}
\begin{enumerate}
\item For $0<\gamma<1$ all roots of $\tilde{g}_s$ have modulus less than 1.
\item At $\gamma=1$, a root of $\tilde{g}_s$ crosses the unit circle at $\lambda=1$ with positive speed; this is a symmetry-breaking bifurcation of the synchronous periodic solution.
\end{enumerate}
\end{prop}

\noindent
{\bf Proof:} \\[-8mm]
\begin{enumerate}
\item For any root $\lambda$ of $\tilde{g}_s$, we have 
\[
   \lambda^n(\lambda-\gamma)=-(1-\gamma) \quad
\ra \quad|\lambda^n(\lambda-\gamma)|=|1-\gamma|. 
\]
Suppose now that $|\lambda|>1$ and $0<\gamma<1$, that is, $1-\gamma > 0$. Then 
\[
  1-\gamma = |\lambda^n(\lambda-\gamma)|  \geq |\lambda|^n(|\lambda|-\gamma) 
   >|\lambda|^n|1-\gamma| 
   > 1-\gamma, 
\]
which is a contradiction.  
\item For $\gamma=1$ we have $\tilde{g}_s(\lambda)=\lambda^n(\lambda-1)=0$ and $\lambda=1$ is a root. By differentiating \eqref{eq:ggtilde} we obtain
\[
 \frac{d\lambda}{d\gamma} 
 = \frac{1+\lambda^n}{\lambda^{n-1}[\lambda(n+1)-n\gamma]}, 
\]
and evaluating this at $\gamma=1,\lambda=1$ we obtain $d\lambda/d\gamma=2 \neq 0$. This is a symmetry-breaking (or pitchfork) bifurcation due to the reflectional symmetry and because this instabilility concerns a direction along which $\theta_1$ and $\theta_2$ deviate from $\theta_1 = \theta_2$; see also Sec.~\ref{sec:synch}\ref{sec:symbrA}. \hfill $\square$
\end{enumerate}

For the synchronous periodic solution to be stable the roots of both $g_s$ and $\tilde{g}_s$ need to have magnitude less than 1. Since $1 <(n+1)/n$, this is the case for $0<\gamma<1$ (as determined by the roots of $\tilde{g}_s$). Moreover, we know from~\cite{laikra22} that $dT/d\tau=0$ when $\gamma=1$, which means that the symmetry-breaking bifurcation, where stability is lost, takes place at the minimum on a curve in the represention in Fig.~\ref{fig:synchper} of the branches of synchronous periodic solutions in terms of $T$ as a function of $\tau$; indeed, as shown, the branches are stable to the right of each minimum in $T$. Each branch loses another stable eigenvalue when a root of $g_s$ passes through 1 at the saddle-node bifurcation at $\gamma=(n+1)/n$; note in Fig.~\ref{fig:synchper} and its inset that this happens at the fold point of each branch with respect to $\tau$. 

The analysis above applies for $n>0$. When $n=0$, $g_s(\lambda)=\lambda-1$ and $\tilde{g}_s(\lambda)=\lambda+1-2\gamma$. Since $\gamma>0$ the only instability that can occur is when $\gamma=1$, corresponding the symmetry breaking bifurcation at the minimum of the black curve shown in Fig.~\ref{fig:synchper}.

\subsection{Bifurcating symmetry-broken synchronous periodic solutions}
\label{sec:symbrA}

To find the periodic solutions that emerge from the points of symmetry breaking on the branches of synchronous periodic solutions in Fig.~\ref{fig:synchper}, we introduce a phase shift $\phi$ and consider 
\[
s_i-t_{i-n}=(n-\phi)T \quad {\rm and} \quad t_i-s_{i-n}=(n+\phi)T,
\]
so that $\phi=0$ corresponds to the perfectly synchronous case. 
Substituting these into~\eqref{eq:spsynch}--\eqref{eq:tpsynch} we obtain the equations \begin{align}
   \coth{((n+1-\phi)T-\tau)} & = \kappa+\coth{((n-\phi)T-\tau)}, \label{eq:existbr} \\
   \coth{((n+1+\phi)T-\tau)} & = \kappa+\coth{((n+\phi)T-\tau)}, \label{eq:existbrA} 
\end{align}
for the existence of this type of periodic solution. Subtracting~\eqref{eq:existbr} from~\eqref{eq:existbrA} and using 
$\coth{x}-\coth{y}=\sinh{(x-y)}/(\sinh{x}\sinh{y})$ gives
\begin{gather}
  \frac{\sinh{(2\phi T)}}{\sinh{((n+1+\phi)T-\tau)}\sinh{((n+1-\phi)T-\tau)}} \nonumber \\
 = \frac{\sinh{(2\phi T)}}{\sinh{((n+\phi)T-\tau)}\sinh{((n-\phi)T-\tau)}} \,. \label{eq:symm}
\end{gather}
Noting that the denominators of~\eqref{eq:symm} must be equal and using
$\sinh{x}\sinh{y}=[\cosh{(x+y)}-\cosh{(x-y)}]/2$, we have
\[
  \cosh{(2(n+1)T-2\tau)} =\cosh{(2nT-2\tau)}.
\]
Since cosh is even, we conclude that one argument must be the negative of the other (since they are not equal), and it follows that the period of the symmtry-broken periodic solutions is $T=2\tau/(2n+1)$. In turn, this means that both~\eqref{eq:existbr} and~\eqref{eq:existbrA} reduce to
\be
  \coth{((1/2-\phi)T)}  = \kappa-\coth{((1/2+\phi)T)} . \label{eq:symbr}
\ee

\begin{figure}[t!]
\begin{center}
\includegraphics[width=10cm]{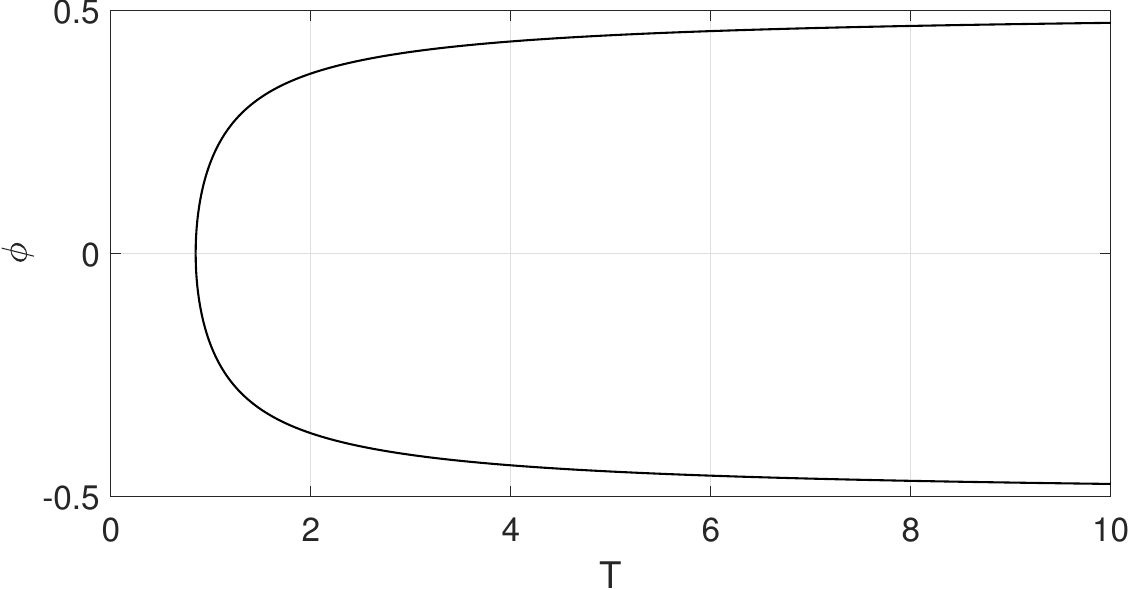}
\caption{The relationship between the phase shift $\phi$ and the period $T$ of the symmetry-broken solutions given by~\eqref{eq:symbr}. Here, $\kappa=5$.}
\label{fig:sols}
\end{center}
\end{figure}

For fixed $\kappa$, solutions of~\eqref{eq:symbr} lie on the curve in $(T,\phi)$ space that is shown in Fig.~\ref{fig:sols} for the specific case $\kappa=5$. There are two branches, one for negative and one for positive $\phi$, which meet at the minimum in $T$ at $2\coth^{-1}{(\kappa/2)}$.  In the $(\tau,T)$ plane of Fig.~\ref{fig:synchper}, the pair of solutions for given $n$ bifurcates from the symmetry-breaking bifurcation points on the $n$th branch of synchronous
solutions; note that both branches project to the segment of the line given by $T = 2\tau/(2n+1)$ for $T$ above the bifurcation point. As we show in  Appendix~\ref{sec:stability_broken}\ref{sec:stabbrokensync}, all these symmetry-broken periodic solutions are unstable.

\section{Alternating solutions}
\label{sec:alt}

\begin{figure}[t!]
\begin{center}
\includegraphics[height=4.2cm]{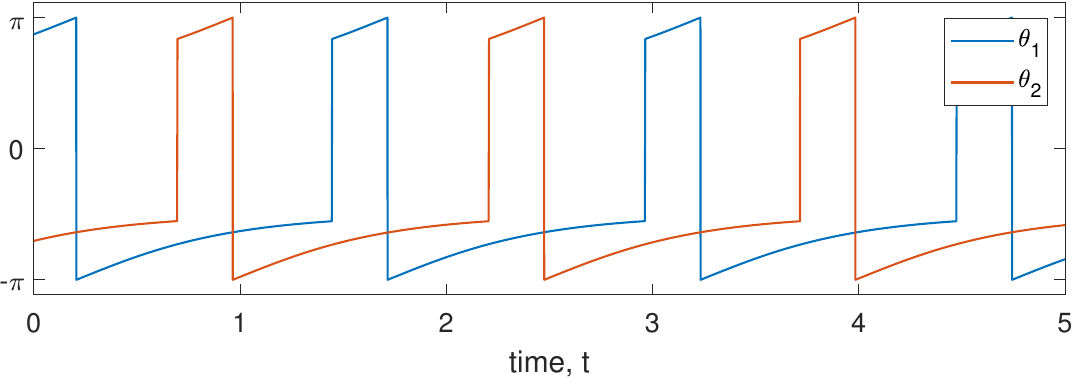}
\caption{Alternating periodic solution of system~\eqref{eq:dth1}--\eqref{eq:dth2} for $\kappa=5$ and $\tau=2$. Note that, when $\theta_i$ reaches $\pi$ from below, it is
reset to $-\pi$.}
\label{fig:alt}
\end{center}
\end{figure}

We now consider solutions for which the neurons take turns firing, and are half a period out of phase with one another. An example of such an alternating solution is shown in Fig.~\ref{fig:alt}; note that, when $\theta$ reaches $\pi$ from below, there is a reset of $\theta$ to $-\pi$.

To determine this type of solution, we consider the general system
\begin{align}
   \frac{dx}{dt} & =  f[x(t),y(t-\tau)] \label{eq:genx}, \\
   \frac{dy}{dt} & =  f[y(t),x(t-\tau)] \label{eq:geny},
\end{align}
and suppose that it has an alternating periodic solution with period $T$, which means $y(t)=x(t+T/2)$ and, hence, $y(t-\tau)=x(t-\tau+T/2)=x(t-(\tau-T/2))$. Thus, this solution is also a periodic solution of period $T$ of the {\em self-coupled} system 
\[
   \frac{dx}{dt} =  f[x(t),x(t-(\tau-T/2))] 
\]
with a delay of $\tau-T/2$. Similarly, if we find a periodic solution with period $T$ of the system
\be
   \frac{dx}{dt}=f(x(t),x(t-\tau)) \label{eq:single}
\ee
(as we did in~\cite{laikra22}), then this solution is also a periodic solution of period $T$ of the system
\begin{align*}
   \frac{dx}{dt} & =  f[x(t),y(t-(\tau+T/2))],  \\
   \frac{dy}{dt} & =  f[y(t),x(t-(\tau+T/2))],
\end{align*}
for which $x$ and $y$ alternate. 

Therefore, since periodic solutions of a single delay-coupled theta neuron of the form~\eqref{eq:single} are given by~\eqref{eq:existSA}, the existence of alternating solutions of~\eqref{eq:dth1}--\eqref{eq:dth2} is given by~\eqref{eq:existSA} when replacing $\tau$ by $\tau-T/2$; that is, the existence of alternating solution is given by the equation
\be
   \coth{[(n+1/2)T-\tau]}=\kappa+\coth{[(n-1/2)T-\tau]}, \label{eq:existSB}
\ee
which is valid for $(n-1/2)T<\tau<(n+1/2)T$. Note that~\eqref{eq:existSB} can also be obtained from~\eqref{eq:existSA} by replacing $n$ by $n-1/2$; however, the physical meaning of this replacement is not clear since $n$ is an integer. The meaning of $n$ in~\eqref{eq:existSB} is that if neuron 1 fires at time $0$, there are $n$ past firing times of neuron 2 in the interval $(-\tau,0)$; note that $n$ could be zero.

\subsection{Stability of alternating solutions}
\label{sec:alternating_stab}

\begin{figure}[t!]
\begin{center}
\includegraphics[width=13.5cm]{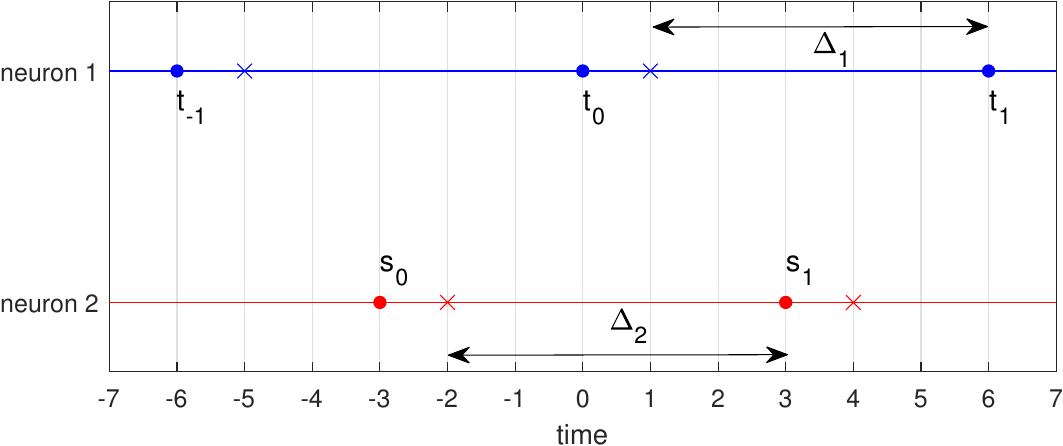}
\caption{Schematic of firing times (circles) and times at which the neuron's phase is
instantaneously increased (crosses). Here the delay is $\tau=4$, the period is $T=6$ and $n=1$; the times $\Delta_{1,2}$ are defined in Sec.~\ref{sec:alt}\ref{sec:alternating_stab}\ref{sec:stab1}.}
\label{fig:schem}
\end{center}
\end{figure}

To determine the stability of alternating solutions we need to consider two cases that depend on the value of the delay $\tau$ relative to the value of the multiple $nT$ of the basic period $T$. While the derivation differs, the resulting equation is actually the same for the two cases. To provide insight into the idea behind either derivation, we show in Fig.~\ref{fig:schem} a schematic of Case 1 described below in Sec.~\ref{sec:alt}\ref{sec:alternating_stab}\ref{sec:stab1}, specifically, for $\tau=4$, $T=6$ and $n=1$. 

\subsubsection{Case 1 where $(n-1/2)T<\tau<nT$} 
\label{sec:stab1}

In Fig.~\ref{fig:schem} with $n=1$, neuron 1 has just fired at time $t=0$. After time $\tau-(t_0-s_{1-n})$, its angle $\theta_1$ equals
\be
   \theta_1^-=2\tan^{-1}{\left[-\coth{\left(\tau-(t_0-s_{1-n})\right)}\right]},
\label{eq:thetamin}
\ee
and this is reset to $\theta_1^+$, where
\[
   \tan{(\theta_1^+/2)}=\tan{(\theta_1^-/2)}+\kappa.
\]
This event is marked with the blue cross at $t=1$ in Fig.~\ref{fig:schem}.
There is then a wait time $\Delta_1$ before neuron 1 fires, which from~\eqref{eq:Delta} is 
\be
   \Delta_1=\coth^{-1}{\left[\tan{\left(\frac{\theta_1^+}{2}\right)}\right]}.
\label{eq:delta1}
\ee
Hence, $t_1=t_0+\tau-(t_0-s_{1-n})+\Delta_1=\tau+s_{1-n}+\coth^{-1}{[\kappa-\coth{(\tau-(t_0-s_{1-n}))}]}$.

Going back to when neuron 2 last fired at $s_0$, in complete analogy, after time $\tau-(s_0-t_{-n})$, we have
\be
   \theta_2^-=2\tan^{-1}{\left[-\coth{\left(\tau-(s_0-t_{-n})\right)}\right]}
\label{eq:theta2min}
\ee
and $\theta_2$ is reset to $\theta_2^+$, where
\[
   \tan{(\theta_2^+/2)}=\tan{(\theta_2^-/2)}+\kappa.
\]
This reset is marked with the red cross at $t=-2$ in Fig.~\ref{fig:schem}. The corresponding wait time $\Delta_2$ before neuron 2 fires is
\be
   \Delta_2=\coth^{-1}{\left[\tan{\left(\frac{\theta_2^+}{2}\right)}\right]}, 
\label{eq:delta2}
\ee
giving  $s_1=s_0+\tau-(s_0-t_{-n})+\Delta_2=\tau+t_{-n}+\coth^{-1}{[\kappa-\coth{(\tau-(s_0-t_{-n}))}]}$.

We calculated $s_1$ and $t_1$ in terms of previous firing times, but the calculation is entirely general, and we conclude for any $i=0,1,2,\dots$ that 
\begin{align}
   s_{i+1} & = \tau+t_{i-n}+\coth^{-1}{[\kappa-\coth{(\tau-s_i+t_{i-n}))}]}, 
\label{eq:sip1A} \\
   t_{i+1} & = \tau+s_{i+1-n}+\coth^{-1}{[\kappa-\coth{(\tau-t_i+s_{i+1-n}))}]}. \label{eq:tip1A}
\end{align}

\subsubsection{Case 2 where $nT<\tau<(n+1/2)T$} 

For this range of $\tau$, suppose that neuron 1 has just fired at $t=t_0=0$ and the previous firing of neuron 2 was at $s_0<t_0$. Then $\theta_2(0)=2\tan^{-1}{\left[-\coth{(t_0-s_0)}\right]}$ and,
after a time $\tau-(t_0-t_{-n})$, the angle $\theta_2$ equals
\begin{align*}
   \theta_2^- & =2\tan^{-1}{\left[-\coth{\left(\tau-(t_0-t_{-n})-\coth^{-1}{\left[\tan{\left(\frac{\theta_2(0)}{2}\right)}\right]}\right)}\right]} \\
  & = 2\tan^{-1}{\left[-\coth{\left(\tau-(s_0-t_{-n})\right)}\right]}.
\end{align*}
This is exactly \eqref{eq:theta2min} and gives the same reset to $\theta_2^+$. Therefore, the wait time $\Delta_2$ before neuron 2 fires is also given by \eqref{eq:delta2}, which yields 
\be
s_1 = t_0+\tau-(t_0-t_{-n})+\Delta_2 
=\tau+t_{-n}+\coth^{-1}{\left[\kappa-\coth^{-1}{(\tau-(s_0-t_{-n}))}\right]}.
\label{eq:case2_s1}
\ee
To find $t_1$, we know that neuron 1 has just fired at $t_0$. In complete analogy, after time $\tau-(t_0-s_{1-n})$, we have $\theta_1=\theta_1^-$ as given by \eqref{eq:theta2min}, with the same reset to $\theta_1^+$ and subsequent wait time $\Delta_1$ given by \eqref{eq:delta1}, yielding
\be
t_1=t_0+\tau-(t_0-s_{1-n})+\Delta_1=\tau+s_{1-n}+\coth^{-1}{[\kappa-\coth{\left(\tau-(t_0-s_{1-n})\right)}]}.
\label{eq:case2_t1}
\ee
As for Case 1, these calculations are again entirely general and, in fact, \eqref{eq:case2_s1} and \eqref{eq:case2_t1} also give equations~\eqref{eq:sip1A}--\eqref{eq:tip1A} for any for $i=0,1,2,\dots$. Note that equation~\eqref{eq:sip1A} is the same expression for $s_{i+1}$ that we found Sec.~\ref{sec:synch}\ref{sec:stab_inphase} for the existence of synchronous solutions; the difference for alternating solutions lies in expression~\eqref{eq:tip1A} for $t_{i+1}$.

\subsubsection{Linearisation around the alternating solution}
\label{sec:linalt}

To linearise around a periodic solution given by~\eqref{eq:sip1A}--\eqref{eq:tip1A}, we write these equations as
\begin{align*}
   R(s_{i+1},t_{i-n},s_i) & = 0, \\
   S(t_{i+1},s_{i-n+1},t_i) & = 0.
\end{align*}
As in Sec.~\ref{sec:synch}\ref{sec:stab_inphase}, we perturb the firing times as $s_i\to s_i+\mu_i$ and $t_i\to t_i+\eta_i$ and obtain to first order
\begin{align*}
  \frac{\partial R}{\partial s_{i+1}}\mu_{i+1}+\frac{\partial R}{\partial t_{i-n}}\eta_{i-n}+\frac{\partial R}{\partial s_{i}}\mu_{i} & =0, \\
  \frac{\partial S}{\partial t_{i+1}}\eta_{i+1}+\frac{\partial S}{\partial s_{i-n+1}}\mu_{i-n+1}+\frac{\partial S}{\partial t_{i}}\eta_{i} & =0,
\end{align*}
which, after calculating the partial derivatives, we write as
\begin{align}
  -\mu_{i+1}+(1-\gamma_1)\eta_{i-n}+\gamma_1\mu_{i} & =0, \label{eq:stA}\\
  -\eta_{i+1}+(1-\gamma_2)\mu_{i-n+1}+\gamma_2\eta_{i} & =0, \label{eq:stB}
\end{align}
where
\begin{align}
   \gamma_1 & =\frac{1-\coth^2{(\tau+t_{i-n}-s_i)}}{1-[\kappa-\coth{(\tau+t_{i-n}-s_i)}]^2}, \label{eq:gam1} \\
   \gamma_2 & =\frac{1-\coth^2{(\tau-t_i+s_{i-n+1})}}{1-[\kappa-\coth{(\tau-t_i+s_{i-n+1})}]^2}. \label{eq:gam2}
\end{align}
For an alternating periodic solution we have
\be
   \gamma := \gamma_1=\gamma_2= \frac{\coth^2{(\tau-(n-1/2)T)}-1}{[\kappa-\coth{(\tau-(n-1/2)T)}]^2-1}. \label{eq:gamma}
\ee
Notice that, when when replacing $n$ by $n+1/2$ in this expression for $\gamma$, we obtain exactly \eqref{eq:gammaA}; in particular, again $\gamma > 0$.

As before for~\eqref{eq:mu}--\eqref{eq:eta}, we consider solutions $\eta_i=A\lambda^i$ and $\mu_i=B\lambda^i$ with $A,B \in \mathbb{R}$ and write the linear difference equations~\eqref{eq:stA}--\eqref{eq:stB} with $\gamma$ as in \eqref{eq:gamma} as the matrix equation
\[
  \begin{pmatrix} (1-\gamma)\lambda^{i-n} & \gamma\lambda^i-\lambda^{i+1} \\
  \gamma\lambda^i-\lambda^{i+1} & (1-\gamma)\lambda^{i-n+1} \end{pmatrix} \begin{pmatrix} A \\ B \end{pmatrix}=\begin{pmatrix} 0 \\ 0 \end{pmatrix}.
\]
Its determinant is
\be (1-\gamma)^2\lambda^{2i-2n+1}-\gamma^2\lambda^{2i}+2\gamma\lambda^{2i+1}-\lambda^{2i+2} \label{eq:detB}
\ee
for given integer $n$. 

The case $n=0$ is special and gives, by setting~\eqref{eq:detB} to zero and then multiplying by $-\lambda^{-2i}$, the characteristic equation
\be
\lambda^2-(1+\gamma^2)\lambda+\gamma^2=(\lambda-1)(\lambda-\gamma^2) =0
\label{eq:stabn=0}
\ee
of the alternating periodic solutions with $n=0$.
We see that $\lambda=1$ is always a solution; it is the trivial Floquet multiplier representing the invariance of periodic solutions under time translation. From the other factor of~\eqref{eq:stabn=0} we conclude that the alternating periodic solution with $n=0$ is stable for 
$0<\gamma<1$ and unstable for $1<\gamma$.

For $n \geq 1$ in~\eqref{eq:detB}, we obtain with multiplication by $-\lambda^{2n-1-2i}$ the characteristic equation
\be 
D_a(\lambda) := \lambda^{2n-1}( \lambda-\gamma)^2-(1-\gamma)^2=0 \label{eq:rootsB}
\ee
of the alternating periodic solutions with $n \geq 1$. We remark that $D_s(\lambda)$ from \eqref{eq:charA} is obtained from $D_a(\lambda)$ by the substitution $n \mapsto n+1/2$. In particular, this characteristic equation also factors, as $D_a(\lambda)=g_a(\lambda)\tilde{g}_a(\lambda)=0$ but now with 
\begin{align*}
   g_a(\lambda) & =\lambda^{n-1/2}(\lambda-\gamma)-(1-\gamma), \\
   \tilde{g}_a(\lambda) & = \lambda^{n-1/2}(\lambda-\gamma)+(1-\gamma).
\end{align*}
These two polynomials in $\rho = \sqrt{\lambda}$, which determine the stability inside the symmetry subspace of the alternating solutions and in the direction transverse to it, respectively, can be treated individually. However, we find it more convenient to consider $D_a(\lambda)$ in \eqref{eq:charA} `holistically' and to factor it via its trivial root $\lambda=1$ as $D_a(\lambda)=(\lambda-1) H_a(\lambda)$, where
\[
H_a(\lambda) = 
\lambda^{2n}+(1-2\gamma)\lambda^{2n-1}+(1-\gamma)^2\sum_{i=0}^{2n-2}\lambda^i .
\]

We have the following statements for the stability of the alternating periodic solutions with $n \geq 1$.

\begin{prop}[Properties of the roots of $D_a$ and $H_a$]
\verb+ + \\[-8mm]  
\label{prop:alternate}
\begin{enumerate}
\item When $\gamma=0$, the roots of $D_a$ are the $(2n+1)$th roots of unity; in particular, they all have modulus 1.
\item When $\gamma=1$, the polyonomial $H_a$ has the single root $1$, as well as the root $0$ with multiplicity $2n-1$ when $n \geq 1$.
\item If $H_a(1)=0$ then either $\gamma=1$ or $\gamma=(2n+1)/(2n-1)$.
\item If $0<\gamma<1$ then no roots of $D_a$ are outside the unit circle, that is, the alternating periodic solution is stable.
\item At both $\gamma=1$ and $\gamma=(2n+1)/(2n-1)$, a real root of $D_a$ transversely leaves the unit circle as $\gamma$ increases.
\item $\gamma=1$ is at the minimum of the graph of $T$ as a function of $\tau$; at this point there is the switch from Case 1 to Case 2 in the derivation of our stability results.
\item At $\gamma=(2n+1)/(2n-1)$ there is a saddle-node bifurcation of alternating periodic solutions.
\item At $\gamma=1$ there is a symmetry bifurcation of alternating periodic solutions.
\end{enumerate}
\end{prop}

\noindent
{\bf Proof:} \\[-8mm]
\begin{enumerate}
\item If $\gamma=0$ then $D_a(\lambda)=\lambda^{2n+1}-1=0$ and the result follows.
\item If $\gamma=1$ then
$H_a(\lambda)=\lambda^{2n}-\lambda^{2n-1}=\lambda^{2n-1}(\lambda-1)=0$
and the result follows.
\item If $H_a(1)=0$ then 
$1+(1-2\gamma)+(1-\gamma)^2(2n-1)=0,$
and solving this quadratic equation for $\gamma$ gives the result.
\item Suppose $D_a(\lambda) = 0$, that is, 
\be
   \lambda^{2n-1}(\lambda-\gamma)^2=(1-\gamma)^2  \ 
   \ra \ |\lambda^{2n-1}(\lambda-\gamma)^2|=(1-\gamma)^2 \label{eq:F1}
\ee
Assume now that $|\lambda|>1$. Then 
$|\lambda^{2n-1}(\lambda-\gamma)^2| > (1-\gamma)^2$, which contradicts~\eqref{eq:F1}; thus, we cannot have $|\lambda|>1$.
\item Differentiating $D_a(\lambda)=0$ with respect to $\gamma$ we get
\[ \frac{d\lambda}{d\gamma}=\frac{2[(\lambda-\gamma)\lambda^{2n-1}+\gamma-1]}{\lambda^{2n-2}[(2n-1)\gamma^2-4n\gamma\lambda+(2n+1)\lambda^2]}.
\]
If $\lambda=1$ and $(2n-1)\gamma^2-4n\gamma+2n+1\neq 0$ (that is, $\gamma\neq1$ and $\gamma\neq(2n+1)/(2n-1)$; compare with (iii))
then $\frac{d\lambda}{d\gamma}=1$, reflecting that the root $\lambda=1$ of $D_a$ is always present. However, if $\gamma=1$ then
\[ \frac{d\lambda}{d\gamma}=\frac{2\lambda(\lambda-1)}{2n-1-4n\lambda+(2n+1)\lambda^2},
\]
which is undefined at $\lambda=1$. With l'Hopital's rule, we conclude that
\[
\lim_{\lambda\to 1}\frac{d\lambda}{d\gamma}=\lim_{\lambda\to 1}\frac{4\lambda-2}{2\lambda(2n+1)-4n}=1>0.
\]
Similarly if $\gamma=(2n+1)/(2n-1)$ then
\[ \frac{d\lambda}{d\gamma}=\frac{2[\lambda(2n-1)-(2n+1)]\lambda^{2n-1}+4}{\lambda^{2n-2}[(2n+1)^2-4n(2n+1)\lambda+(4n^2-1)\lambda^2]},
\]
which is also undefined at $\lambda=1$. Again, with l'Hopital's rule we have
\[
\lim_{\lambda\to 1}\frac{d\lambda}{d\gamma}=\lim_{\lambda\to 1}\frac{\lambda (2 n - 1)(2 n \lambda - 2 n - 1)}{4 n^3 (\lambda - 1)^2 - n [(\lambda - 2) \lambda + 3] - 1}=\frac{2n-1}{2n+1}>0,
\]
where we used that $n \geq 1$.
\item Since~\eqref{eq:existSB} defines $T$ as a function of $\tau$, we can differentiate it with respect to $\tau$ to get
\be \csch^2{[(n+1/2)T-\tau]}\left[(n+1/2)\frac{dT}{d\tau}-1\right]=\csch^2{[(n-1/2)T-\tau]}\left[(n-1/2)\frac{dT}{d\tau}-1\right] \label{eq:diff}
\ee
Setting $dT/d\tau=0$ we get (by taking reciprocals)
\[
  \coth^2{[(n+1/2)T-\tau]}=\coth^2{[(n-1/2)T-\tau]}. 
\]
We cannot have 
\[
   \coth{[(n+1/2)T-\tau]}=\coth{[(n-1/2)T-\tau]}
\]
because that would imply from~\eqref{eq:existSB}
that $\kappa=0$, which is not possible, so we conclude that
\be
   \coth{[(n+1/2)T-\tau]}=-\coth{[(n-1/2)T-\tau]}. \label{eq:6A}
\ee 
Substituting~\eqref{eq:6A} into~\eqref{eq:existSB} we obtain 
\[
   \kappa=-2\coth{[(n-1/2)T-\tau]}=2\coth{[\tau-(n-1/2)T]},
\]
and substituting this into~\eqref{eq:gamma} we find $\gamma=1$, as required.

From~\eqref{eq:6A} we have $(n+1/2)T-\tau=\tau-(n-1/2)T$ which can be rearranged to the condition $\tau=nT$ for switching between Cases 1 and 2. 
\item We first write $\gamma$ from~\eqref{eq:gamma} as 
\begin{align}
   \gamma & =  \frac{\coth^2{[(n-1/2)T-\tau]}-1}{\{\kappa+\coth{[(n-1/2)T-\tau]}\}^2-1} \nonumber \\
& = \frac{\coth^2{[(n-1/2)T-\tau]}-1}{\coth^2{[(n+1/2)T-\tau]}-1} \nonumber \\
& = \frac{\csch^2{[(n-1/2)T-\tau]}}{\csch^2{[(n+1/2)T-\tau]}}, \label{eq:7A}
\end{align}
where we used~\eqref{eq:existSB} to get the second line. We also rearrange~\eqref{eq:diff} to obtain
\be \frac{dT}{d\tau}=\frac{\csch^2{[(n+1/2)T-\tau]}-\csch^2{[(n-1/2)T-\tau]}}{\csch^2{[(n+1/2)T-\tau]}(n+1/2)-\csch^2{[(n-1/2)T-\tau]}(n-1/2)} . \label{eq:dTdtau}
\ee
Now, if $\gamma=(2n+1)/(2n-1)$ then, from~\eqref{eq:7A},
\[
   (2n-1)\csch^2{[(n-1/2)T-\tau]}=(2n+1)\csch^2{[(n+1/2)T-\tau]} ,
\]
and, thus, the denominator of~\eqref{eq:dTdtau} is zero. However, the numerator of~\eqref{eq:dTdtau} is not zero (since $\gamma\neq 1$); hence, the graph of $T$ as a function of $\tau$ has an infinite slope. This identifies the change of stability at $\gamma=(2n+1)/(2n-1)$ from (v) as a saddle-node bifurcation. 
\item Since we have a minimum of $T$ as a function of $\tau$ at $\gamma=1$, with derivative zero, the change of stability according to (v) at this point is not a saddle-node bifurcation. In light of the symmetry of system~\eqref{eq:dth1}--\eqref{eq:dth2}, this is indeed a symmetry-breaking bifurcation. \hfill $\square$
\end{enumerate}

\subsection{Branches of alternating periodic solutions}
\label{sec:alternate_branches}

\begin{figure}[t!]
\begin{center}
\includegraphics[height=10.2cm]{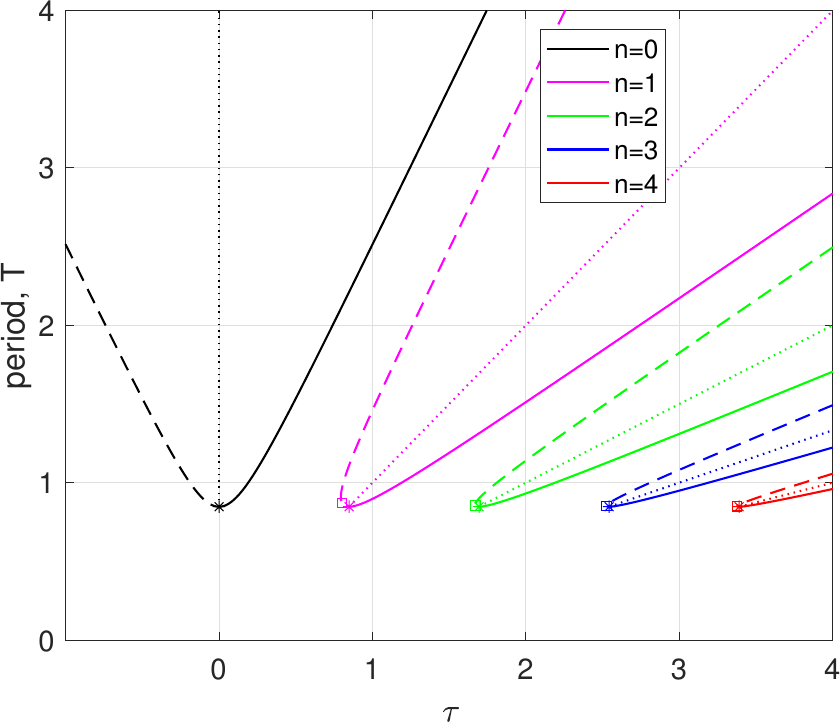}
\caption{Branches of alternating periodic solutions of system~\eqref{eq:dth1}--\eqref{eq:dth2} with $n=0,1,2,3,4$. Solid curves are stable (to the right of each minimum) and dashed curves are unstable. Saddle-node bifurcations are marked with squares, symmetry-breaking bifurcations with stars, and bifurcating pairs of symmetry-broken solutions are represented by dotted lines. Here, $\kappa=5$.}
\label{fig:altper}
\end{center} 
\end{figure}

The branches of alternating solutions given by~\eqref{eq:existSB} are shown in Fig.~\ref{fig:altper} for $\kappa = 5$ and $n$ up to $4$, with stability information as given by Proposition~\ref{prop:alternate}. The branch for $n=0$ plays a special role, and we describe it first. This branch of alternating solutions is symmetric with respect to reflection about the $T$-axis, and it has the minumum $\overline{T}=2\coth^{-1}{(\kappa/2)}$ at $\tau = 0$. Moreover, $0<\gamma<1$ for $\tau>0$ and $1<\gamma$ for $\tau<0$. This means that the alternating solutions with  $n=0$ and $\tau > 0$ are stable; however, those with $\tau < 0$ are not. From the minimum, where $\gamma=1$, a vertical branch of symmetry-broken alternating solutions for $\tau=0$ emerges; they are discussed below in Sec.~\ref{sec:alt}\ref{sec:symbrB}. The stable branch for $n=0$ is interesting since it persists down to $\tau=0$, that is, such an alternating solution exists even without delays. This can be understood as follows. Suppose neuron 1 fires. This immediately pushes neuron 2 past its threshold but, since it takes a finite time for neuron 2 to then fire, the influence of neuron 2 firing is not immediately felt by neuron 1. Due to this reaction time, neuron 2 effectively provides delayed feedback of neuron 1 to itself (and vice versa, by swapping neuron labels).  

All branches with $n \geq 1$ also have the minimum $\overline{T}=2\coth^{-1}{(\kappa/2)}$, where the stability changes, as well as a point of saddle-node bifurcation. As stated in Proposition~\ref{prop:alternate}, $\gamma$ varies monotonically along each of these branches, and it is between 0 and 1 to the right of the minimum; hence, these alternating solutions, which are described by Case 1, are stable. Each minimum is a symmetry-breaking bifurcation, and a branch of symmetry-broken alternating solutions emerges. These branches are already shown in Fig.~\ref{fig:altper}, and they are discussed in Sec.~\ref{sec:alt}\ref{sec:symbrB}. The alternating solutions to the other side of the minimum along any branch with $n \geq 1$ are described by Case 2 and unstable. Continuing along a branch, a second multiplier leaves the unit circle through 1 at the saddle-node bifurcation where $\gamma=(2n+1)/(2n-1)$. Along the respective upper branch, the alternating periodic solutions have two unstable directions.

\subsection{Bifurcating symmetry-broken alternating periodic solutions}
\label{sec:symbrB}

\begin{figure}[t!]
\begin{center}
\includegraphics[width=11cm]{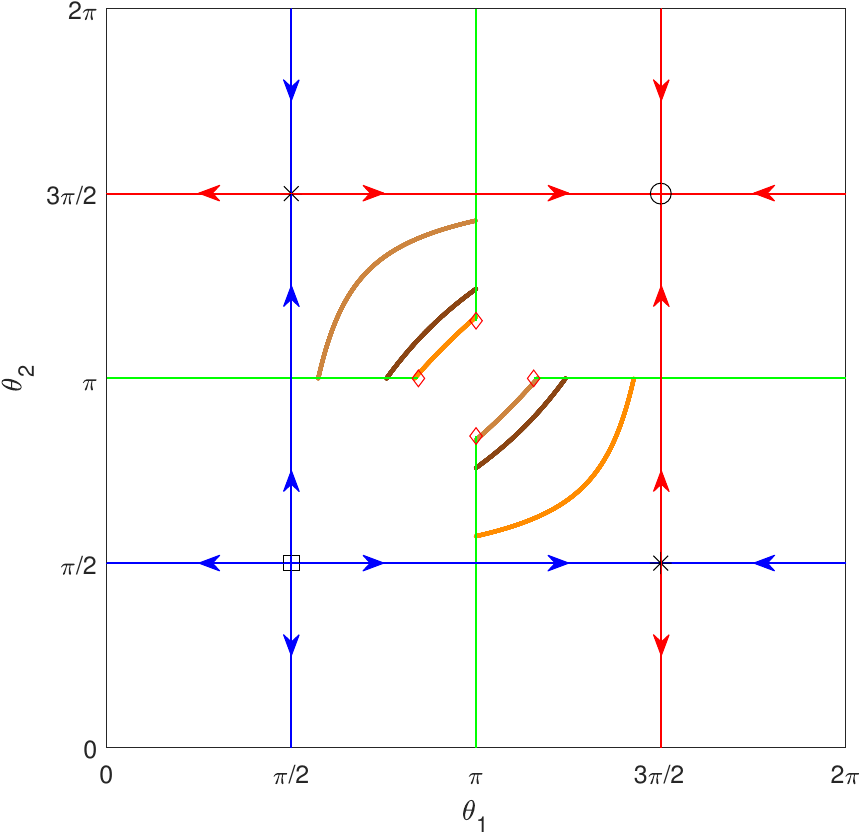}
\caption{Illustration of alternating periodic solutions of system~\eqref{eq:dth1}--\eqref{eq:dth2} with $\tau=0$ as trajectories of system~\eqref{eq:nodelA}--\eqref{eq:nodelB} on the square $[0, 2\pi] \times [0, 2\pi]$ with the firing lines (green) where one phase increases through $\pi$ and a discontinuous reset occurs. Shown is a pair of symmetry-broken periodic solutions (light brown and orange curves), which pass through $(\theta_1,\theta_2)=(1.6,\pi)$ and $(\theta_1,\theta_2)=(2\pi - 1.6,\pi)$, respectively, and the symmetric periodic solution at the point of symmetry breaking (brown curves), which has the minimum period $\overline{T}=2\coth^{-1}{(\kappa/2)}$ and passes through  $(\theta_1,\theta_2)=(2\tan^{-1}{(\kappa/2)},\pi)$ and $(\theta_1,\theta_2)=(2\pi - 2\tan^{-1}{(\kappa/2)},\pi)$. Also shown are the repellor (square), the attractor (circle), the two saddle equilibria (crosses) and their stable and unstable manifolds (blue and red horizontal and vertical lines). The red diamonds bound the ranges of $\theta_1$ and $\theta_2$ along which any alternating periodic solutions with $\tau=0$ must fire. Here, $\kappa=5$.}
\label{fig:nodelay}
\end{center}
\end{figure}

We first consider the symmetry-broken alternating periodic solutions for the special case that $n=0$, which bifurcate from the minimum at $\tau=0$ and exist along the vertical dashed line in Fig~\ref{fig:altper}. Since there is no delay,
system~\eqref{eq:dth1}--\eqref{eq:dth2} can be visualised in the phase space $\mathbb{S}^1 \times \mathbb{S}^1$ of the periodic variables $\theta_1$ and $\theta_1$, which we represent by the square $[0, 2\pi] \times [0, 2\pi]$ subject to identification of its left and right, and lower and upper boundaries. The firing with reset of the two neurons occurs along the lines where $\theta_1 = \pi$ and $\theta_2 = \pi$. Between firings, the flow is given by the system of the two ordinary differential equations (ODEs)
\begin{eqnarray}
   \frac{d\theta_1}{dt} & = & -2\cos{\theta_1} \label{eq:nodelA}, \\
   \frac{d\theta_2}{dt} & = & -2\cos{\theta_2} \label{eq:nodelB},
\end{eqnarray} 
which has four equilibria: the source $(\pi/2,\pi/2)$, the sink $(3\pi/2,3\pi/2)$, and the saddles $(\pi/2,3\pi/2)$ and $(3\pi/2,\pi/2)$. As is shown in Fig.~\ref{fig:nodelay}, these equilibria are connected by the stable and unstable invariant manifolds of the saddles, which are vertical and horizontal lines because \eqref{eq:nodelA} and \eqref{eq:nodelB} are decoupled. Note the symmetry of Fig.~\ref{fig:nodelay} with respect to reflection in the diagonal, due to the exchange symmetry $(\theta_1,\theta_2) \mapsto (\theta_2,\theta_1)$.

To find periodic solutions of system~\eqref{eq:nodelA}--\eqref{eq:nodelB}, we assume that $\theta_2$ has just fired, that is, $\theta_2=\pi$ and $\theta_1=\alpha$. As is clear from Fig.~\ref{fig:nodelay}, when $0 \leq \alpha< \pi/2$ or $3\pi/2 \leq \alpha \leq 2\pi$ the trajectory will approach the attractor $(\theta_1,\theta_2)=(3\pi/2,3\pi/2)$. For $\alpha=\pi/2$ the trajectory approaches the saddle equilibrium $(\theta_1,\theta_2)=(\pi/2,3\pi/2)$ since this initial point lies on its stable manifold. However, when $\pi/2<\alpha<\pi$ 
both $\theta_1$ and $\theta_2$ increase until the firing line with $\theta_1=\pi$ is reached at $\theta_2=2\tan^{-1}{[-\coth{(\Delta_1)}]}=2\pi-\alpha$ after the time
$\Delta_1=\coth^{-1}{[\tan{(\alpha/2)}]}$. At this point, $\theta_2$ is reset and incremented to $\theta_2^+=2\tan^{-1}{(\kappa+\tan{(\pi-\alpha/2)})}$, which we assume is greater than $\pi/2$ (otherwise the system approaches an equilibrium). Both phases then continue to increase until $\theta_2=\pi$, which takes a further time $\Delta_2=\coth^{-1}{[\tan{(\theta_2^+/2)}]}$, at which point $\theta_1=2\tan^{-1}{[-\coth{(\Delta_2)}]}=2\pi-\theta_2^+$. Now $\theta_1$ is reset and incremented to $\theta_1^+=2\tan^{-1}{(\kappa+\tan{(\pi-\theta_2^+/2)})}$. For this process to yield a periodic solution we need $\theta_1^+=\alpha$, which is, in fact, true for any $\alpha$ that results in a return to this firing line. This is due to the special form of system~\eqref{eq:nodelA}--\eqref{eq:nodelB} and the symmetry of $\cos$ around $\pi$: the angle $\theta_1$ increases from $\alpha$ to $\pi$ by the same amount as $\theta_2$, which therefore increases to $2\pi-\alpha$; the equivalent statement is true for the second `leg' of the periodic solution with $\theta_1>\pi$. 

Hence, there is a one-parameter family of (pairs of) symmetry-broken periodic solutions. The pair for $\alpha = 1.6$ and $\alpha= 2\pi - 1.6$ is shown in Fig.~\ref{fig:nodelay}; also shown is the symmetric alternating periodic solution for $\alpha= 2\tan^{-1}{(\kappa/2)}$ with period $\overline{T}=2\coth^{-1}{(\kappa/2)}$, from which the family of symmetry-broken periodic solutions bifurcate. More generally, the existence of this family of periodic solutions is due to the fact that system~\eqref{eq:nodelA}--\eqref{eq:nodelB} is reversible; specifically, it is invariant under the transformation $(t,\theta_1,\theta_2)\mapsto(-t,-\theta_1,-\theta_2)$, which is the rotation by $\pi$ about the origin (and any $\pi$ translates) with time reversal. In light of the exchange symmetry, the reversibility manifests itself in Fig.~\ref{fig:nodelay} as the reflection in the antidiagonal, subject to the reversal of time $t$. 

The condition that $\pi/2<\theta_2^+$ in order to obtain a periodic solution is equivalent to $\tan{(\alpha/2)}<\kappa-1$ and, thus, restricts the valid range for $\alpha$ to the intervals $(\pi/2,\alpha^\ast)$ and $(\pi/2,2\pi - \alpha^\ast)$ with $\alpha^\ast=2\tan^{-1}{(\kappa-1)}$.  The corresponding points $(\alpha^\ast,\pi)$ and $(2\pi - \alpha^\ast,\pi)$ on the $\theta_2$-firing line, as well as $(\pi,\alpha^\ast)$ and $(\pi,2\pi - \alpha^\ast)$ on the $\theta_1$-firing line, represent the `inner' limit of the family of symmetry broken alternating solutions; these four points are marked in Fig.~\ref{fig:nodelay}. The other limit of this family for $\alpha \searrow \pi/2$ and for $\alpha \nearrow 3\pi/2$ is formed by the respective parts of the invariant manifolds of the two saddles. Note that $\alpha^\ast\to\pi/2$ as $\kappa \searrow 2$, and for $\kappa\leq 2$ alternating solutions with $\tau=0$ do not exist. 

The period of any solution of the family is $T=\Delta_1+\Delta_2$, and we can write $\Delta_1=(1/2+\phi)T$ and $\Delta_2=(1/2-\phi)T$ for some $-1/2<\phi<1/2$. We find that $\coth{(\Delta_1)}=\tan{(\alpha/2)}$ and $\coth{(\Delta_2)}=\kappa-\tan{(\alpha/2)}$ and, thus, $\coth{(\Delta_2)}=\kappa-\coth{(\Delta_1)}$, or
\be
  \coth{((1/2-\phi)T)}  = \kappa-\coth{((1/2+\phi)T)}, \label{eq:symbrT}
\ee
which is exactly~\eqref{eq:symbr}. Hence, the graph of the period of the symmetry broken alternating solutions as a function of $\phi$ is that shown in Fig.~\ref{fig:sols}, whose minimum period $\overline{T}=2\coth^{-1}{(\kappa/2)}$ occurs when $\Delta_1=\Delta_2$, that is, when $\alpha=2\tan^{-1}{(\kappa/2)}$. As we show in Appendix~\ref{sec:stability_broken}\ref{sec:stabbrokenaltzero}, the periodic solutions with $\tau=0$ are neutrally stable, which is consistent with there being a continuum of them.

To find the branches of symmetry-broken alternating periodic solutions for $n \geq 1$ with $\tau >0$, we use the concept of reappearance \cite{yanper09}: a periodic solution with $\tau=0$ with a given $\phi$ and $T$ satisfying~\eqref{eq:symbrT} is also a periodic solution with the same $\phi$ and $T$ when the delay equals an integer multiple $nT$ of the period $T$. These symmetry-broken solutions, hence, lie on the straight line segments $T=\tau/n$ with $T>2\coth^{-1}{(\kappa/2)}$, which emerge from the minima, the points of symmetry-breaking bifurcations, on the branch of alternating periodic solutions for the respective $n \geq 1$. All these symmetry-broken solutions are unstable, as we show in Appendix~~\ref{sec:stability_broken}\ref{sec:stabbrokenaltpos}.

\subsection{Overall picture of periodic solutions}
\label{sec:soverall}

\begin{figure}[t!]
\begin{center}
\includegraphics[width=11.9cm]{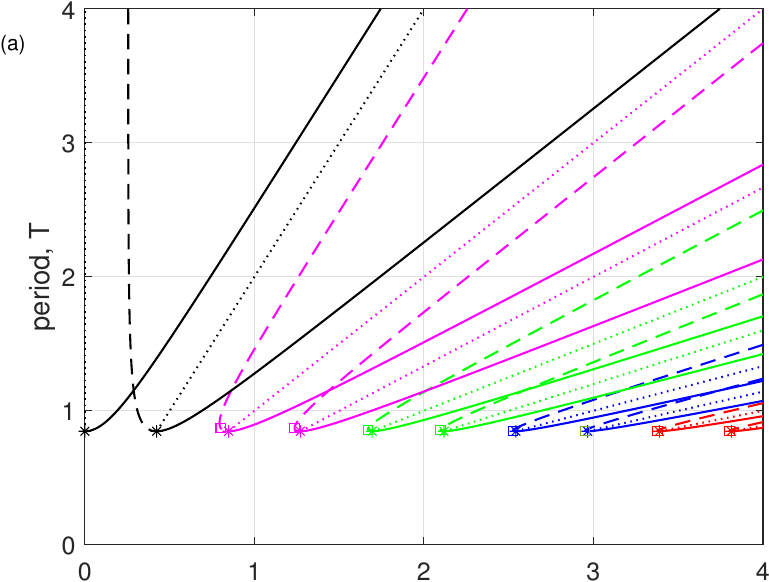} \\[5mm]
\includegraphics[width=11.9cm]{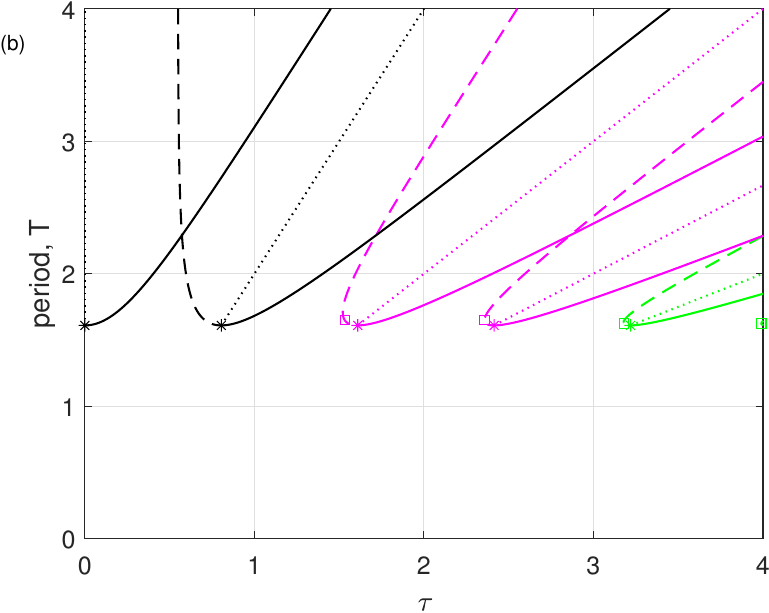} 
\caption{All solution branches found in Secs.~\ref{sec:synch} and~\ref{sec:alt} plotted together, for $\kappa=5$ in panel~(a) and for $\kappa=3$ in panel~(b). Symbols and colours for each $n$ are as in Figs.~\ref{fig:synchper} and~\ref{fig:altper}; the leftmost curve of the same colour is that of alternating solutions and the rightmost one of synchronous solutions.}
\label{fig:both}
\end{center}
\end{figure}

We finish our investigation by presenting together in Fig.~\ref{fig:both} the branches of synchronous and alternating periodic solutions and the respective branches of symmtery-broken solutions; here, panel~(a) shows the combined solution branches for $\kappa=5$ from Figs.~\ref{fig:synchper} and~\ref{fig:altper}, and panel~(b) shows the equivalent image for $\kappa=3$. Fig.~\ref{fig:both} shows that the two types of periodic solution branches interleave, with symmetry-broken solutions of one type fitting into the `gaps' between branches of the other type. Notice that the `neighbouring' curves forming these gaps have the limiting slopes $T=\tau/n$ of the respective line segment of symmetry-broken solutions. As we observed earlier, the consecutive branches (with larger and larger $\tau$-values of their minima) map to one another when the transformation $n \mapsto n + 1/2$ is applied to their respective parametrisation, which entails a switch of type; note that this mapping is consistent with the stability properties: the branches of synchronised and of alternating periodic solutions are qualitatively the same in this regard.

We concentrated on investigating the effects of varying the delay $\tau$, but it is also of interest to vary the coupling strength $\kappa$. For all of the periodic solutions we found, $\kappa$ must be greater than 2. This is the result of setting $I=-1$ at the start of the paper, as the Dirac delta function coupling must be strong enough to kick a neuron across the `gap' between the stable and unstable equilibria of the theta neuron (whose position is related to $I$) in order to trigger its firing. As $\kappa$ is decreased towards $\kappa = 2$, the distance between the branches increases, but note that the slopes of the branches of symmetry-broken solutions are independent of $\kappa$. In the limit $\kappa \searrow 2$, all branches with minima disappear to the right, towards infinitely large values of $\tau$, and the stable branch of alternating solutions with $n=0$ is the only one left in the region with $\tau>0$. 

When $\kappa$ is increased, on the other hand, all symmetry-breaking and saddle-node bifurcation points approach the origin of the $(\tau,T)$-plane. Specifically, the minimum period $T=2\coth^{-1}{(\kappa/2)}$ of solution branches in Fig.~\ref{fig:both} goes to zero, and the `gaps' between the stable part of the $n$th and the unstable part of the $(n+1)$th branch of the same type shrink to zero as well. The dynamics are considerably simplified in the limit $\kappa\to\infty$. From~\eqref{eq:reset} we see that, in this case, the influence of a past firing of a neuron is to cause the receiving neuron to immediately fire. From~\eqref{eq:existSA}, synchronous periodic solution therefore exist for $T=\tau/(n+1)$. From~\eqref{eq:gammaA} we have $\gamma=0$ and, thus, from~\eqref{eq:charA} these solutions are neutrally stable. Similar arguments show that alternating periodic solutions exist for $T=\tau/(n+1/2)$ and are also neutrally stable. Moreover, no symmetry-broken solutions exist. Hence, the limit $\kappa \to \infty$ in the $(\tau,T)$-plane is an infinite fan of interleaving rays from the origin, of synchronous and alternating periodic solutions, with the rays accumulating on the $\tau$-axis.

\section{Two theta neurons coupled with smooth feedback}
\label{sec:smooth}

\begin{figure}
\begin{center}
\includegraphics[width=14cm]{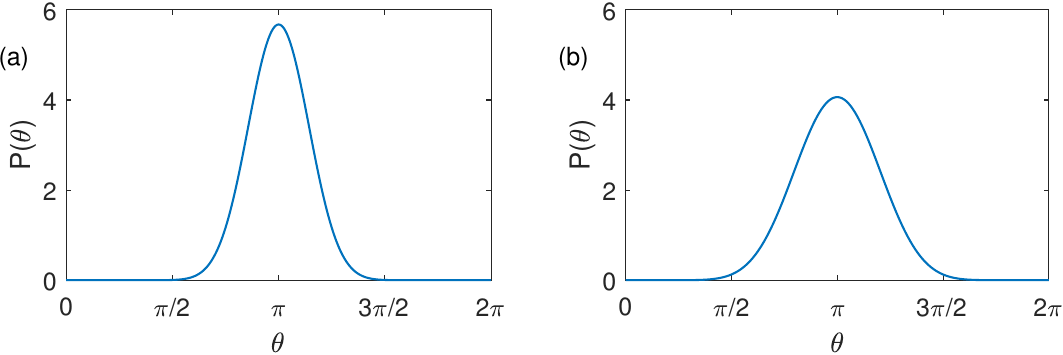} 
\caption{The pulsatile function $P(\theta)$ for $m = 10$ in panel~(a) and $m = 5$ in panel~(b).}
\label{fig:pulseplot}
\end{center}
\end{figure}

We now consider the case of smooth feedback with the pulsatile function centred at $\theta=\pi$ given by 
\be
   P(\theta)=a_m(1-\cos{\theta})^m, \label{eq:bumP}
\ee
where $a_m=2^m(m!)^2/(2m)!$ ensures that $\int_0^{2\pi}P(\theta)d\theta=2\pi$
for all $m$; see also \cite{laikra22}. Increasing $m$ makes this function `sharper', and in the limit $m\to\infty$ we have $P(\theta)=2\pi\delta(\theta-\pi)$ where $\delta$ is the Dirac delta function. With this smooth feedback, the equations of the two coupled theta neurons are
\begin{align}
   \frac{d\theta_1}{dt} & =1-\cos{\theta_1}+(1+\cos{\theta_1})\left(-1+\kappa P[\theta_2(t-\tau)]\right), \label{eq:sm1} \\
    \frac{d\theta_2}{dt} & =1-\cos{\theta_2}+(1+\cos{\theta_2})\left(-1+\kappa P[\theta_1(t-\tau)]\right), \label{eq:sm2}
\end{align}
which is a delay differential equation (DDE) with the constant delay $\tau$ that has system~\eqref{eq:dth1}--\eqref{eq:dth2} as its limit for $m\to\infty$ (although with a 
different value of $\kappa$).

We proceed by discussing the branches of its different types of periodic solutions, as well as their stability properties; this information is no longer available analytically, but it can be found with the continuation package DDE-BIFTOOL \cite{NewDDEBiftool} . Specifically, we consider $P(\theta)$ for the two cases $m = 10$ and $m = 5$ that are shown in Fig.~\ref{fig:pulseplot}.

\subsection{Synchronised solutions}

\begin{figure}[t!]
\begin{center}
\includegraphics[height=9.6cm]{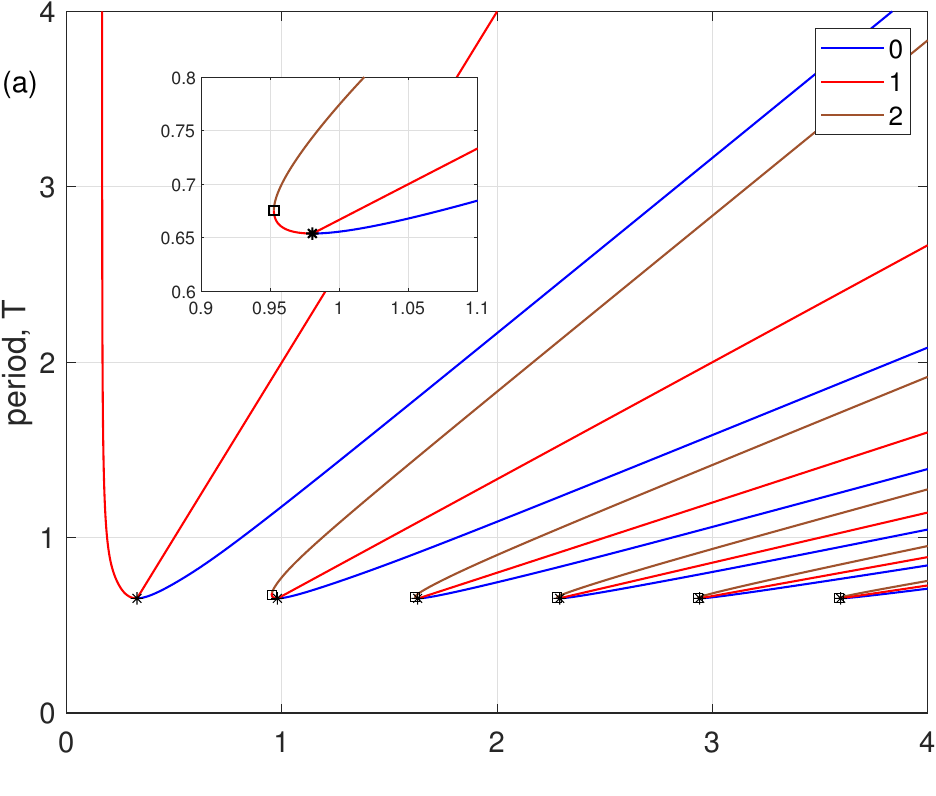} \\[1.2mm]
\includegraphics[height=9.6cm]{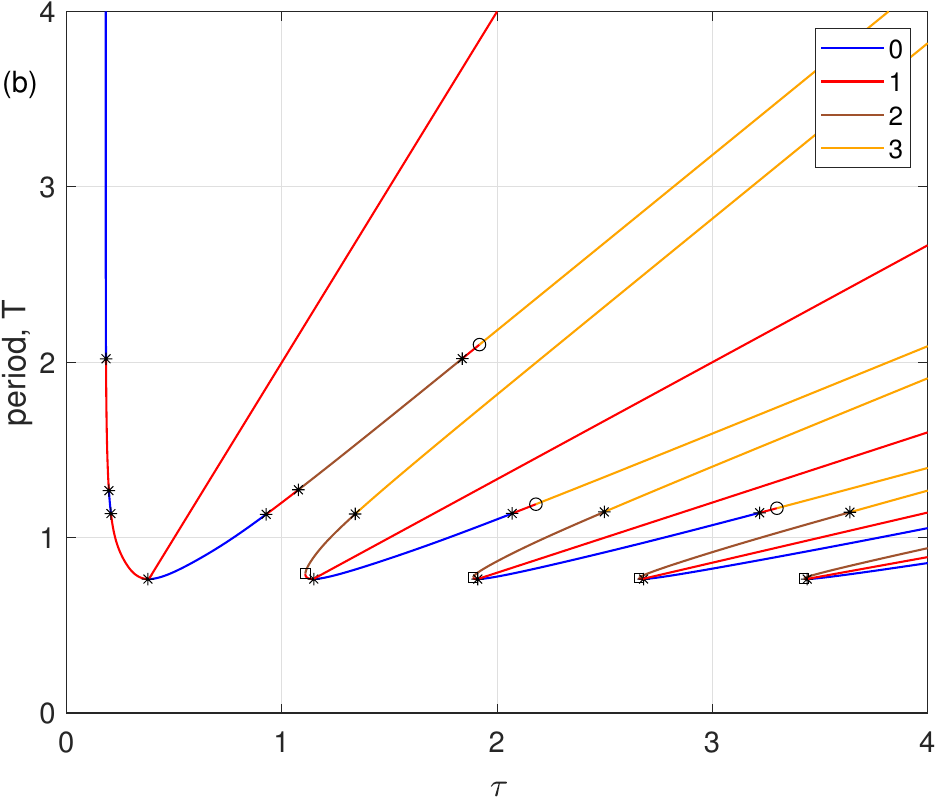} 
\caption{Branches of synchronised and corresponding symmetry-broken periodic solutions of~\eqref{eq:sm1}--\eqref{eq:sm2} with $\kappa=5$ for $m=10$ in panel~(a) and $m = 5$ in panel~(b). Colour indicates the number
of unstable Floquet multipliers, capped at three; saddle-node bifurcations are marked with squares, symmetry- breaking bifurcations with stars, and Hopf bifurcations with circles. The inset of panel~(a) is an enlargement of the second branch near the saddle-node and symmetry- breaking bifurcations. Compare with Fig.~\ref{fig:synchper}.}
\label{fig:smoothsync}
\end{center}
\end{figure}

Fig.~\ref{fig:smoothsync} shows the branches of synchronised periodic solutions of~\eqref{eq:sm1}--\eqref{eq:sm2} with $\kappa=5$ and for $m=10$ in panel~(a) and $m = 3$ in panel~(b). They we obtained by continuing the respective stable periodic solution in the delay $\tau$, while monitoring the Floquet multipliers to detect bifurcations due to stability changes. The symmetry-broken synchronised periodic solutions were continued from the symmetry-breaking points by making use of the branch switching capability of DDE-BIFTOOL. The branches of the different periodic solutions are very close to and qualitatively the same as those in Fig.~\ref{fig:synchper}(a). In particular, as for the limiting case of a Dirac delta function coupling, the branches in Fig.~\ref{fig:smoothsync} emerge for increasing $\tau$ at saddle-node bifurcations and there is a symmetry-breaking bifurcation at the minimum of each branch, with a stable part of the branch to its right. Note that the branches of symmetry-broken periodic solutions still lie exactly on the lines $T=2\tau/(2n+1)$, indicating that this is a property of the solutions rather than the specific form of the model. 

These statements are true for both $m=10$ and $m = 5$, but there is a considerable difference in the observed stability along branches. For the reasonably sharp pulsatile function $P(\theta)$ with $m=10$ in Fig.~\ref{fig:smoothsync}(a), even the stability properties agree with those for the limiting case in Fig.~\ref{fig:synchper}. In other words, the analytical results, including those on the stability of branches, give the correct description of the properties of the synchronised periodic solutions over the ranges of $\tau$ and $T$ that are shown here. When $P(\theta)$ is much less sharp, as in Fig.~\ref{fig:smoothsync}(b) for $m=5$, we find quite a number of additional stability changes, including Hopf bifurcations. While the stability of the branches agrees with that for~\eqref{eq:dth1}--\eqref{eq:dth2} sufficiently close to the saddle-node and symmetry-breaking bifurcations, more Floquet multipliers of the synchronised periodic solutions leave or re-enter the unit circle as $\tau$ is increased. This observation agrees with that in~\cite{laikra22} for the case of a single neuron with delayed self-feedback. For the symmetry-broken synchronised periodic solutions, on the other hand, we do not find additional stability changes.

\subsection{Alternating solutions}

\begin{figure}[t!]
\begin{center}
\includegraphics[height=9.6cm]{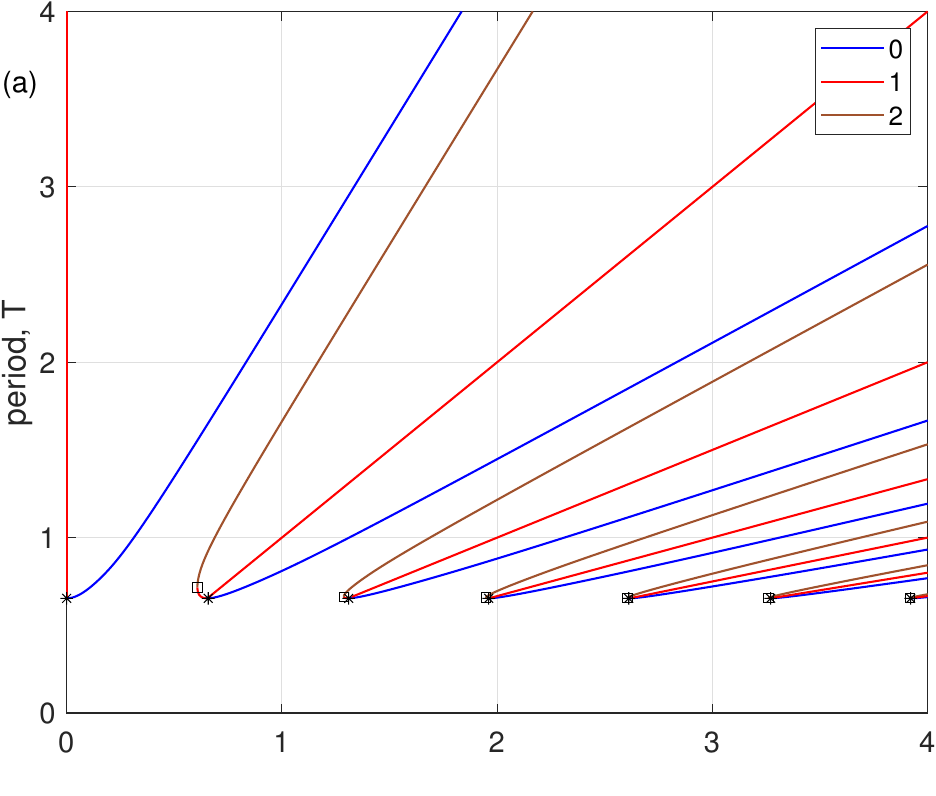} \\[1.2mm]
\includegraphics[height=9.6cm]{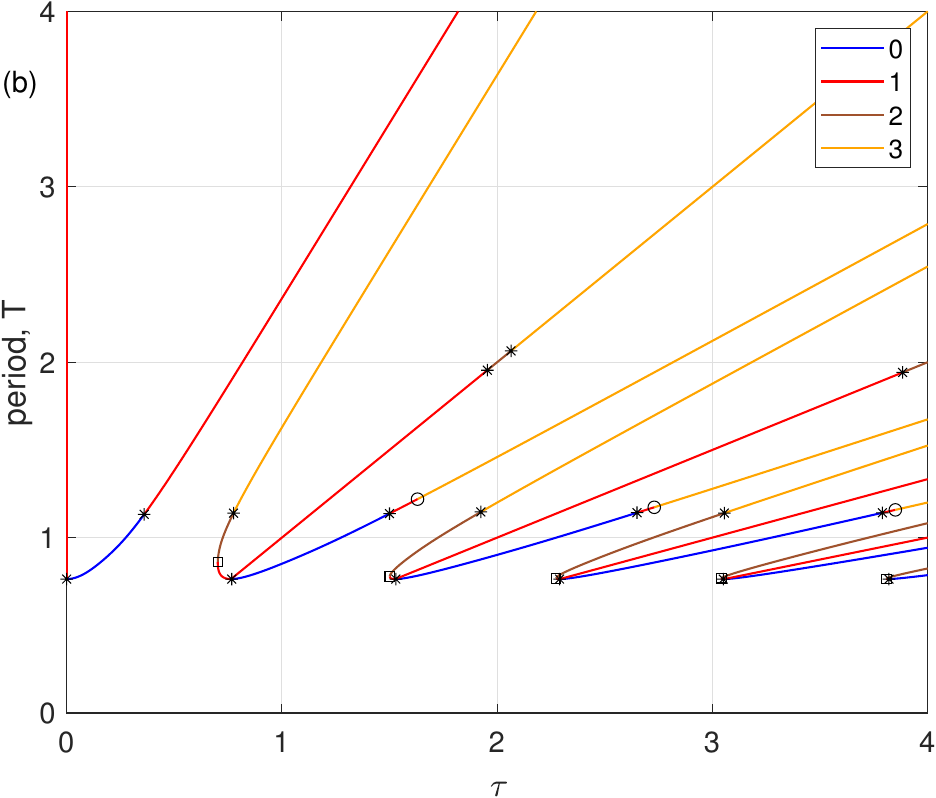} 
\caption{Branches of alternating and corresponding symmetry-broken periodic solutions of~\eqref{eq:sm1}--\eqref{eq:sm2} with $\kappa=5$ for $m=10$ in panel~(a) and $m = 5$ in panel~ (b), shown as in Fig.~\ref{fig:smoothsync}. Compare with Fig.~\ref{fig:altper}.}
\label{fig:smoothalt}
\end{center}
\end{figure}

As expected, system~\eqref{eq:sm1}--\eqref{eq:sm2} also supports alternating periodic solutions of the type studied here; their computed branches are shown in Fig.~\ref{fig:smoothalt}, again with $\kappa=5$ and for $m=10$ in panel~(a) and $m = 5$ in panel~(b). Also for this type of solution we find that these branches are very close to and qualitatively similar to the corresponding ones in Fig.~\ref{fig:altper} for the Dirac delta function coupling. Moreover, the stability is still maintained over the same ranges for the `sharp case' of $P(\theta)$ with $m=10$, while there are again additional changes of the Floquet multipliers of the alternating periodic solutions when $m = 5$. Importantly, in both cases, except for the branch starting from $\tau=0$, the branches are born in saddle-node bifurcations and feature subsequent symmetry breaking at their minima.

\subsubsection{Symmetry-broken alternating solutions with $\tau=0$}

\begin{figure}[t!]
\begin{center}
\includegraphics[width=11.2cm]{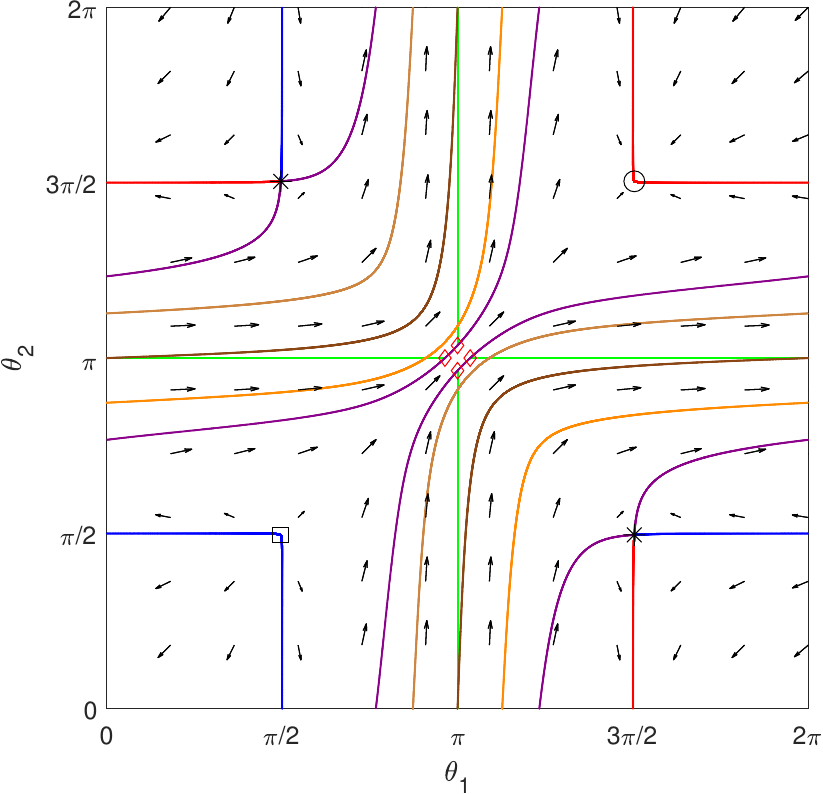}
\caption{Phase portrait on the square $[0, 2\pi] \times [0, 2\pi]$ of the planar ODE given by sytem~\eqref{eq:sm1}--\eqref{eq:sm2} with $\tau=0$, showing: the repellor (square), the attractor (circle), the two saddle equilibria (crosses) and their invariant manifolds (blue, red, and purple curves when forming a homoclinic connection), a pair of symmetry-broken periodic solutions (light brown and orange curves) of the one-parameter family, and the limiting neutrally stable symmetry-broken periodic solution (brown curve). Also shown are the firing lines (green), which are bounded by their intersection points with the homoclinic connections (red diamonds). Here, $\kappa=5$ and $m=10$, and the arrows indicate the flow. Compare with Fig.~\ref{fig:nodelay}.}
\label{fig:smoothtauz}
\end{center}
\end{figure}

For $\tau=0$ the smooth system~\eqref{eq:sm1}--\eqref{eq:sm2} reduces to a planar system of ODEs, which can be characterised completely. As was the case for the limiting system~\eqref{eq:nodelA}--\eqref{eq:nodelB}, this smooth equivalent is invariant under both the interchange of neurons 1 and 2, and the reversibility transformation $(t,\theta_1,\theta_2)\mapsto(-t,-\theta_1,-\theta_2)$; the latter is due to the fact that the pulsatile function $P(\theta)$ from \eqref{eq:bumP} is symmetric about $\theta=\pi$ and about $\theta=0$. The phase portrait for $\kappa=5$ and $m=10$ is shown in Fig.~\ref{fig:smoothtauz}. It features the same equilibria as system~\eqref{eq:nodelA}--\eqref{eq:nodelB}, but the stable and unstable manifolds of the two saddles $(\theta_1,\theta_2) = (\pi/2,3\pi/2)$ and $(\theta_1,\theta_2) = (3\pi/2,\pi/2)$ are now not straight lines. While one branch comes from the repellor, or goes to the attractor, the other branches from a pair of homoclinic connections, one to each saddle; compare with Fig.~\ref{fig:nodelay}. As Fig.~\ref{fig:smoothtauz} shows, a region with a family of (neutrally stable) periodic solutions on the torus is bounded by these two homoclinic connections, which also bound the firing lines where, when crossed in the positive direction, the respective neuron fires. A pair of periodic solutions emerges from the unique neutrally stable alternating periodic solutions in Fig.~\ref{fig:smoothtauz}. One such pair is shown, and the period of these periodic solutions increases without bound as they approach the homoclinic connections. Trajectories that are not periodic end up at an equilibrium, namely the attractor $(\theta_1,\theta_2) = (3\pi/2,3\pi/2)$ for typical initial conditions. We remark that this mix of `conservative' and `dissipative' dynamics in different parts of the phase space is a common property of reversible systems~\cite{robqui92}; in particular, it has been discussed previously for a network of three coupled theta neurons~\cite{lai18A}.

\section{Discussion and outook}
\label{sec:conclusions}

We studied perfectly synchronous as well as alternating periodic solutions
in a pair of delay-coupled excitatory theta neurons with coupling via a Dirac
delta function, given by system~\eqref{eq:dth1}--\eqref{eq:dth2}. This system acts as a `normal form' for the more general situation of two excitable systems that are mutually coupled subject to a delay --- in that it allowed us to determine the existence and stability of these periodic solutions analytically and explicitly. We were also able to explicitly determine the existence and stability of symmetry-broken periodic solutions that bifurcate from the synchronous and alternating ones. 

This work can be seen as a natural extension of the case of a single theta neurons with Dirac delta function self-feedback, which we studied in~\cite{laikra22}. In fact, the perfectly synchronous solutions can be viewed as arising from self-coupling, and their existence is given by the same expression as for the latter. However, synchronous solutions have different stability properties: they become unstable at the minimum period as a function of the delay $\tau$, rather than at a saddle-node bifurcation as was the case for a single neuron~\cite{laikra22}. This change of stability is a symmetry-breaking bifurcation, creating periodic solutions for which the two neurons no longer fire at the same time. Similarly, we considered alternating periodic solution and showed that they exist with the same period as a perfectly synchronous one, but at a value of $\tau$ that is  increased (or decreased) by half a period compared to that of the perfectly synchronous solution; the stability properties of these two types of periodic solutions are qualitatively the same; in particular, the minima of the period are again symmetry-breaking bifurcations, from which symmetry-broken alternating periodic solutions emerge for which the two neurons no longer fire exactly half a period out of phase with one another. These symmetry-broken solutions exist and can be analysed when $\tau=0$, and branches for $\tau>0$ exist due to the reappearance of periodic solutions in delay differential equations~\cite{yanper09}.  

Overall, there is a clear relationship between the two types of solutions --- branches of synchronous periodic solutions given by~\eqref{eq:existSA} and alternating periodic solutions given by~\eqref{eq:existSB}. They are interleaved for an increasing number $(n+1)$ of spikes per period, and can be obtained from one another by the transformation $n\mapsto n+1/2$. In terms of the delay, this transformation corresponds to replacing $\tau$ by $\tau+T/2$, which can be interpreted as `generated' by the spatio-temporal symmetry of alternating periodic solutions. This principle applies to the stability calculations as well: by applying the transformation $n\mapsto n+1/2$, the characteristic equation~\eqref{eq:rootsB} for the alternating solution is identified as that of the synchronous solution in~\eqref{eq:charA}; the same is true for the characteristic equations~\eqref{eq:H1} and~\eqref{eq:H5} of the respective  symmetry-broken periodic solutions.

As a first confirmation that the system is representative in the spirit of a normal form, we also considered system~\eqref{eq:sm1}--\eqref{eq:sm2} of two theta neurons with smooth, non-impulsive coupling. The resulting smooth DDE was investigated by means of numerical continuation, and this showed that the structure of synchronous and alternating periodic solutions, as well as the respective symmetry-broken ones, is qualitatively the same. Moreover, the stability of these periodic solutions is the same even for coupling via a quite wide pulse; however, when the pulse becomes too wide, there may be additional changes of stability along branches of periodic solutions.

An interesting phenomenon in delay-coupled systems is a phase-flip~\cite{prakur06,adhpra11}, where the phase between oscillators jumps from 0 to $\pi$ as the delay is varied. This cannot happen in system~\eqref{eq:dth1}--\eqref{eq:dth2} with Dirac delta function coupling, since a stable branch remains stable as $\tau$ is increased. However, a phase-flip may be possible when $\tau$ is decreased, or when the coupling is non-instantaneous --- as in system~\eqref{eq:sm1}--\eqref{eq:sm2}, where we indeed found additonal stability changes for increasing $\tau$ for the widest case of the coupling pulse we considered.

\begin{figure}[t!]
\begin{center}
\includegraphics[height=4.3cm]{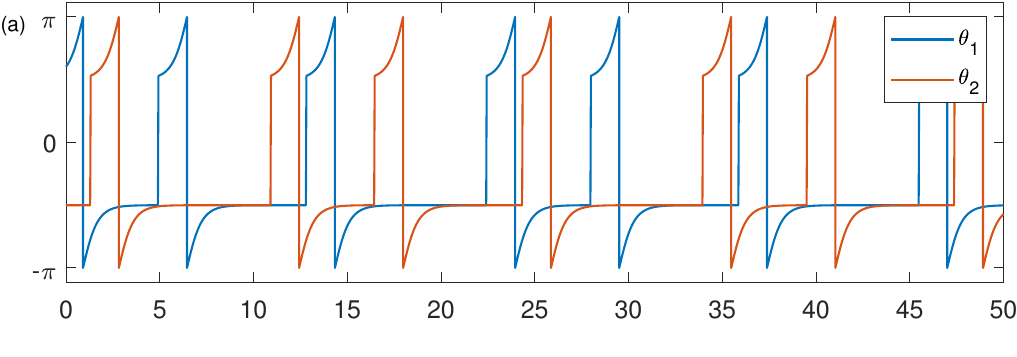} \\
\includegraphics[height=4.3cm]{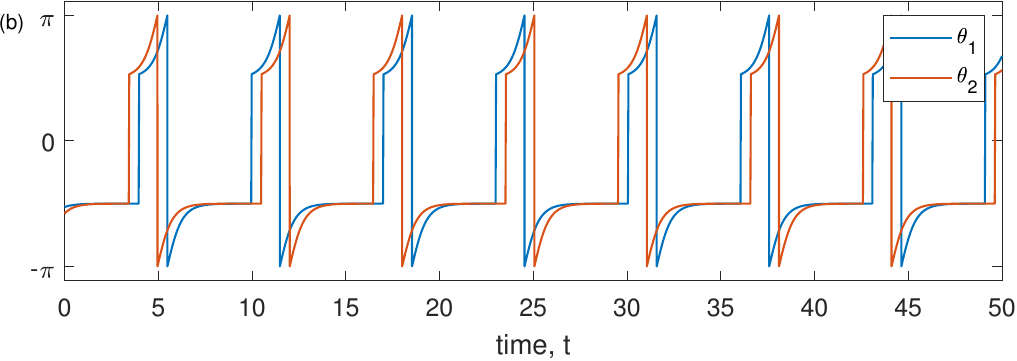}
\caption{Two stable periodic solutions of system~\eqref{eq:dth1}--\eqref{eq:dth2} of different types; here, $\kappa=2.1$ with $\tau=10$ in panel~(a) and $\tau=5$ in panel~(b).  Compare with Fig.~\ref{fig:alt}.}
\label{fig:other}
\end{center}
\end{figure}

We focused here on synchronous and alternating periodic solutions, but there also exist more complicated periodic solutions in system~\eqref{eq:dth1}--\eqref{eq:dth2}, and they may be stable. Figure~\ref{fig:other} shows two examples that we found by numerical integration. In panel~(a) the two neurons alternate their firing times; however, during the period of approximately 23, each neuron fires three times before the pattern repeats. In Fig.~\ref{fig:other}(b) each neuron fires twice within the period of about 12, but without an intervening firing of the other neuron. We remark that a periodic solution with such structure was observed in simulations in~\cite{horgen12}. These different types of periodic solutions could be analysed with the methods presented here, although the more firing times there are within the period, the more difficult it is to keep track of them in the calculations required for existence and stability.

While here we concentrated on the excitable case, with $I=-1$, a similar analysis can be done for the free-running case, with $I=1$ without loss of generality, of two regularly firing theta neurons that are delay-coupled with a Dirac delta function. Two branches of periodic solutions are found, one corresponding to synchronous solutions and the other to alternating solutions. The two branches are connected by symmetry-broken solutions which arise when the period of a solution as a function of delay is at a maximum or a minimum. These results are presented in~\cite{lai24}.

Possible generalisations in the same spirit could include investigating periodic solutions of an excitatory-inhibitory pair of neurons, which may support a PING rhythm~\cite{borkop05}, or of more complicated networks of three or more theta neurons with Dirac delta function coupling, such as a ring of unidirectionally coupled neurons~\cite{kliluc18,peryan10}.

\appendix
\section{Instability of symmetry-broken periodic solutions}
\label{sec:stability_broken}

We show here that the symmetry-broken periodic solutions are all unstable, first for those of the synchronous and then for those of the alternating case. The key aspect is that we now have two parameters, $\gamma_1$ and $\gamma_2$, which are different.

\subsection{Synchronous case}
\label{sec:stabbrokensync}

To find the stability of the symmetry-broken solutions found in Sec.~\ref{sec:synch}\ref{sec:symbrA}, we consider~\eqref{eq:Fpartial}--\eqref{eq:Gpartial} and evaluate partial derivatives to obtain~\eqref{eq:stA}--\eqref{eq:stB} as in Sec.~\ref{sec:alt}\ref{sec:alternating_stab}\ref{sec:linalt}, but now with
\begin{align} \gamma_1 & =\frac{\coth^2{(\tau-s_i+t_{i-n})}-1}{[\kappa-\coth{(\tau-s_i+t_{i-n})}]^2-1}
=\frac{\coth^2{(\tau-(n-\phi)T)}-1}{[\kappa-\coth{(\tau-(n-\phi)T)}]^2-1}, 
\label{eq:g1sync}\\
  \gamma_2 & =\frac{\coth^2{(\tau-t_i+s_{i-n})}-1}{[\kappa-\coth{(\tau-t_i+s_{i-n})}]^2-1}= \frac{\coth^2{(\tau-(n+\phi)T)}-1}{[\kappa-\coth{(\tau-(n+\phi)T)}]^2-1}
\label{eq:g2sync}
\end{align}
for the symmetry-broken periodic solution with given $\phi$. Again, with $\eta_i=A\lambda^i$ and $\mu_i=B\lambda^i$, \eqref{eq:stA}--\eqref{eq:stB} with $\gamma_1$ and $\gamma_2$ from~\eqref{eq:g1sync}--\eqref{eq:g2sync} can be written as
\be
  \begin{pmatrix} (1-\gamma_1)\lambda^{i-n} & \gamma_1\lambda^i-\lambda^{i+1} \\
  \gamma_2\lambda^i-\lambda^{i+1} & (1-\gamma_2)\lambda^{i-n} \end{pmatrix} \begin{pmatrix} A \\ B \end{pmatrix}=\begin{pmatrix} 0 \\ 0 \end{pmatrix},
\label{eq:matrixstab}
\ee
yielding the characteristic equation
\be
   \widehat{D}_s(\lambda) := \lambda^{2n}(\lambda-\gamma_1)(\lambda-\gamma_2) - (1-\gamma_1)(1-\gamma_2)=0 \label{eq:Dhat_s}
\ee
for the symmetry-broken synchronous solutions (which gives $D_s(\lambda)$ from \eqref{eq:charA} for $\gamma_1= \gamma_2$). Substituting $T=2\tau/(2n+1)$ into the definitions of $\gamma_1$ and $\gamma_2$, and using~\eqref{eq:symbr}, one can show that
 $\gamma_1\gamma_2=1$  and $\beta := \gamma_1+\gamma_2>2$. 
Thus,
\[ \widehat{D}_s(\lambda)=\lambda^{2n}(\lambda^2-\beta\lambda+1)+\beta-2, 
\]
which factorises as $\widehat{D}_s(\lambda)=(\lambda-1)\widehat{H}_s(\lambda)$ where 
\be \widehat{H}_s(\lambda)= 
\lambda^{2n+1} +(1-\beta) \lambda^{2n} +(2-\beta)\sum_{i=0}^{2n-1}\lambda^i. \label{eq:H1}
\ee
Since $\widehat{H}_s(1)=(2n+1)(2-\beta)<0$ and $\lim_{\lambda\to\infty} \widehat{H}_s(\lambda)=\infty$, the polynomial $\widehat{H}_s(\lambda)$ has at least one real root greater than 1; thus the symmetry-broken synchronous solutions found in Sec.~\ref{sec:synch}\ref{sec:symbrA} and shown in Fig.~\ref{fig:synchper} are unstable.

\subsection{Alternating case for $\tau=0$}
\label{sec:stabbrokenaltzero}

The firing times~\eqref{eq:sip1A}--\eqref{eq:tip1A} of the alternating periodic solutions with $n=0$ and $\tau=0$ are 
\begin{align}
   s_{i+1} & = t_{i}+\coth^{-1}{[\kappa+\coth{(s_i-t_{i})}]} \label{eq:sip2}, \\
   t_{i+1} & = s_{i+1}+\coth^{-1}{[\kappa+\coth{(t_i-s_{i+1})}]} \label{eq:tip2},
\end{align}
and we write them as
\begin{align*}
   R(s_{i+1},t_{i},s_i) & = 0, \\
   S(t_{i+1},s_{i+1},t_i) & = 0.
\end{align*}
Perturbing the firing times as $t_i\to t_i+\eta_i$ and $s_i\to s_i+\mu_i$ and calculating the partial derivatives, we again obtain \eqref{eq:stA}--\eqref{eq:stB}, but now with 
\begin{align*}
   \gamma_1 & =\frac{1-\coth^2{(t_{i}-s_i)}}{1-[\kappa-\coth{(t_{i}-s_i)}]^2}=\frac{1-\coth^2{((1-\phi)T)}}{1-[\kappa-\coth{((1-\phi)T)}]^2},  \\
   \gamma_2 & =\frac{1-\coth^2{(-t_i+s_{i+1})}}{1-[\kappa-\coth{(-t_i+s_{i+1})}]^2}=\frac{1-\coth^2{(\phi T)}}{1-[\kappa-\coth{(\phi T)}]^2}.
\end{align*}
Assuming again that $\eta_i=A\lambda^i$ and $\mu_i=B\lambda^i$ for some constants $A,B$, we obtain the matrix equation 
\[
  \begin{pmatrix} (1-\gamma_1)\lambda^{i} & \gamma_1\lambda^i-\lambda^{i+1} \\
  \gamma_2\lambda^i-\lambda^{i+1} & (1-\gamma_2)\lambda^{i+1} \end{pmatrix} \begin{pmatrix} A \\ B \end{pmatrix}=\begin{pmatrix} 0 \\ 0 \end{pmatrix}, 
\]
which gives, after multiplication by $-\lambda^{-2i}$ of its determinant, the characteristic equation
\be
\lambda^{2}-(1+\gamma_1\gamma_2)\lambda+\gamma_1\gamma_2
=(\lambda-1)(\lambda-\gamma_1\gamma_2) =0 
\label{eq:roots}
\ee
for the symmetry-broken alternating solutions with $n=0$. The first factor gives the trivial Floquet multiplier $\lambda=1$, and the second factor a second Floquet multiplier $\lambda=1$ because, again, $\gamma_1\gamma_2=1$, since~\eqref{eq:symbrT} also hold in this case. Thus, these solutions are neutrally stable.

\subsection{Alternating case for $\tau>0$}
\label{sec:stabbrokenaltpos}

The firing times are given by~\eqref{eq:sip1A}--\eqref{eq:tip1A}, and we again linearise around a periodic solution to obtain~\eqref{eq:stA}--\eqref{eq:stB} with
\begin{align*}
   \gamma_1 &
=\frac{1-\coth^2{(\tau-(n-1/2+\phi)T)}}{1-[\kappa-\coth{(\tau-(n-1/2+\phi)T)}]^2}, \\
   \gamma_2 & 
=\frac{1-\coth^2{(\tau-(n-1/2-\phi) T)}}{1-[\kappa-\coth{(\tau-(n-1/2-\phi) T)}]^2}
\end{align*}
Assuming again that $\eta_i=A\lambda^i$ and $\mu_i=B\lambda^i$ for some constants $A,B$, we obtain
 \[
   \begin{pmatrix} (1-\gamma_1)\lambda^{i-n} & \gamma_1\lambda^i-\lambda^{i+1} \\
   \gamma_2\lambda^i-\lambda^{i+1} & (1-\gamma_2)\lambda^{i-n+1} \end{pmatrix} \begin{pmatrix} A \\ B \end{pmatrix}=\begin{pmatrix} 0 \\ 0 \end{pmatrix}.
 \]
Multiplying the determinant of the matrix above by $-\lambda^{2n-1-2i}$ gives the characteristic equation
\[ \widehat{D}_a(\lambda) := \lambda^{2n-1}(\lambda-\gamma_1)(\lambda-\gamma_2) - (1-\gamma_1)(1-\gamma_2)=0 
\]
for the symmetry-broken alternating solutions with $n \geq 1$. Note that $\widehat{D}_s$ from \eqref{eq:Dhat_s} is obtained from $\widehat{D}_a(\lambda)$ by the substitution $n \mapsto n+1/2$. With~\eqref{eq:symbrT}, and the fact that these solutions have period $T=\tau/n$, it follows again that $\gamma_1\gamma_2=1$ and $\beta = \gamma_1+\gamma_2>2$. Paralleling the argument in 
Appendix~\ref{sec:stability_broken}\ref{sec:stabbrokensync}, we have 
\[
     \widehat{D}_a(\lambda)=\lambda^{2n-1}(\lambda^2-\beta\lambda+1) +\beta-2 =(\lambda-1)\widehat{H}_a(\lambda)
\]
with
\be 
\widehat{H}_a(\lambda)=\lambda^{2n} +(1-\beta) \lambda^{2n-1} + (2-\beta)\sum_{i=0}^{2n-2}\lambda^i, \label{eq:H5}
\ee
which, when substituting $n \mapsto n+1/2$ here, is exactly $\widehat{H}_s(\lambda)$ from \eqref{eq:H1}. Hence, by the same argument  and with $n \geq 1$ here, $\widehat{H}_a(\lambda)$ has at least one real root greater then 1, and the symmetry-broken alternating solutions found in Sec.~\ref{sec:alt}\ref{sec:symbrB} and shown in Fig.~\ref{fig:altper} are also unstable.

\bibliographystyle{plain}
\bibliography{delTheta_references}

\begin{thebibliography}{10}

\bibitem{adhpra11}
Bhim~Mani Adhikari, Awadhesh Prasad, and Mukeshwar Dhamala.
\newblock Time-delay-induced phase-transition to synchrony in coupled bursting
  neurons.
\newblock {\em Chaos: An Interdisciplinary Journal of Nonlinear Science},
  21(2):023116, 2011.

\bibitem{borkop05}
Christoph B{\"o}rgers and Nancy Kopell.
\newblock Effects of noisy drive on rhythms in networks of excitatory and
  inhibitory neurons.
\newblock {\em Neural computation}, 17(3):557--608, 2005.

\bibitem{burtod03}
Nikola Buri\ifmmode~\acute{c}\else \'{c}\fi{} and Dragana
  Todorovi\ifmmode~\acute{c}\else \'{c}\fi{}.
\newblock Dynamics of fitzhugh-nagumo excitable systems with delayed coupling.
\newblock {\em Phys. Rev. E}, 67:066222, Jun 2003.

\bibitem{dahhil09}
Markus~A Dahlem, G~Hiller, Anastasiia Panchuk, and Eckehard Sch{\"o}ll.
\newblock Dynamics of delay-coupled excitable neural systems.
\newblock {\em International Journal of Bifurcation and Chaos},
  19(02):745--753, 2009.

\bibitem{dodsen04}
Ramana Dodla, Abhijit Sen, and George~L. Johnston.
\newblock Phase-locked patterns and amplitude death in a ring of delay-coupled
  limit cycle oscillators.
\newblock {\em Phys. Rev. E}, 69:056217, May 2004.

\bibitem{erm96}
Bard Ermentrout.
\newblock {Type I} membranes, phase resetting curves, and synchrony.
\newblock {\em Neural Computation}, 8(5):979--1001, 1996.

\bibitem{ermkop86}
G~B Ermentrout and N~Kopell.
\newblock Parabolic bursting in an excitable system coupled with a slow
  oscillation.
\newblock {\em SIAM J Appl Math}, 46(2):233--253, 1986.

\bibitem{GarbinNC15}
Bruno Garbin, Julien Javaloyes, Giovanna Tissoni, and St{\'e}phane Barland.
\newblock Topological solitons as addressable phase bits in a driven laser.
\newblock {\em Nature Communications}, 6, jan 2015.

\bibitem{golshaefer}
Martin Golubitsky, Ian Stewart, and David~G Schaeffer.
\newblock {\em Singularities and Groups in Bifurcation Theory: Volume II},
  volume~69.
\newblock Springer Science \& Business Media, 2012.

\bibitem{horgen12}
Viktor Horvath, Pier~Luigi Gentili, Vladimir~K Vanag, and Irving~R Epstein.
\newblock Pulse-coupled chemical oscillators with time delay.
\newblock {\em Angewandte Chemie}, 124(28):6984--6987, 2012.

\bibitem{IzhikevichBook}
E.M. Izhikevich.
\newblock {\em Dynamical Systems in Neuroscience: The Geometry of Excitability
  and Bursting.}
\newblock The MIT press, 2007.

\bibitem{KelleherPRE10}
B.~Kelleher, C.~Bonatto, P.~Skoda, S.~P. Hegarty, and G.~Huyet.
\newblock Excitation regeneration in delay-coupled oscillators.
\newblock {\em Physical Review E}, 81(3), mar 2010.

\bibitem{kliluc18}
Vladimir Klinshov, Leonhard L{\"u}cken, Serhiy Yanchuk, and Vladimir Nekorkin.
\newblock Multi-jittering instability in oscillatory systems with pulse
  coupling.
\newblock In Mark Edelman, Elbert E.~N. Macau, and Miguel A.~F. Sanjuan,
  editors, {\em Chaotic, Fractional, and Complex Dynamics: New Insights and
  Perspectives}, pages 261--285. Springer, 2018.

\bibitem{klinek11}
VV~Klinshov and VI~Nekorkin.
\newblock Synchronization of time-delay coupled pulse oscillators.
\newblock {\em Chaos Solitons Fractals}, 44(1-3):98--107, 2011.

\bibitem{exciteoverview}
B.~Krauskopf, K.~R. Schneider, J.~Sieber, S.~M. Wieczorek, and M.~Wolfrum.
\newblock Excitability and self-pulsations near homoclinic bifurcations in
  semiconductor laser systems.
\newblock {\em Optics Communications}, 215:367--379, 2003.

\bibitem{KrauskopfWalker}
B.~Krauskopf and J.~J. Walker.
\newblock Bifurcation study of a semiconductor laser with saturable absorber
  and delayed optical feedback.
\newblock In Kathy L{\"u}dge, editor, {\em Nonlinear Laser Dynamics}, pages
  161--181. Wiley-VCH, 2012.

\bibitem{lai18A}
Carlo~R Laing.
\newblock Chaos in small networks of theta neurons.
\newblock {\em Chaos: An Interdisciplinary Journal of Nonlinear Science},
  28(7):073101, 2018.

\bibitem{lai24}
Carlo~R. Laing.
\newblock Periodic solutions for a pair of delay-coupled active theta neurons.
\newblock {\em to appear in The ANZIAM Journal}, 2024.

\bibitem{laikra22}
Carlo~R Laing and Bernd Krauskopf.
\newblock Theta neuron subject to delayed feedback: a prototypical model for
  self-sustained pulsing.
\newblock {\em Proc. R. Soc. A}, 478:20220292, 2022.

\bibitem{lailon03}
Carlo~R Laing and Andr{\'e} Longtin.
\newblock Dynamics of deterministic and stochastic paired excitatory-inhibitory
  delayed feedback.
\newblock {\em Neural computation}, 15(12):2779--2822, 2003.

\bibitem{lilin17}
P.~Li, W.~Lin, and K.~Efstathiou.
\newblock Isochronous dynamics in pulse coupled oscillator networks with delay.
\newblock {\em Chaos}, 27(5):053103, 2017.

\bibitem{peryan10}
P.~Perlikowski, S.~Yanchuk, O.~V. Popovych, and P.~A. Tass.
\newblock Periodic patterns in a ring of delay-coupled oscillators.
\newblock {\em Phys. Rev. E}, 82:036208, Sep 2010.

\bibitem{pophau06}
Oleksandr~V Popovych, Christian Hauptmann, and Peter~A Tass.
\newblock Control of neuronal synchrony by nonlinear delayed feedback.
\newblock {\em Biological Cybernetics}, 95(1):69--85, 2006.

\bibitem{prakur06}
Awadhesh Prasad, J{\"u}rgen Kurths, Syamal~Kumar Dana, and Ramakrishna
  Ramaswamy.
\newblock Phase-flip bifurcation induced by time delay.
\newblock {\em Physical Review E}, 74(3):035204, 2006.

\bibitem{robqui92}
John~AG Roberts and GRW Quispel.
\newblock Chaos and time-reversal symmetry. order and chaos in reversible
  dynamical systems.
\newblock {\em Physics Reports}, 216(2-3):63--177, 1992.

\bibitem{RomeiraNSR16}
B.~Romeira, R.~Av\'o, J.~M.~L. Figueiredo, S.~Barland, and J.~Javaloyes.
\newblock Regenerative memory in time-delayed neuromorphic photonic resonators.
\newblock {\em Scientific Reports}, 6, 2016.

\bibitem{schhil09}
Eckehard Sch{\"o}ll, Gerald Hiller, Philipp H{\"o}vel, and Markus~A Dahlem.
\newblock Time-delayed feedback in neurosystems.
\newblock {\em Philosophical Transactions of the Royal Society A: Mathematical,
  Physical and Engineering Sciences}, 367(1891):1079--1096, 2009.

\bibitem{schwag89}
Heinz~Georg Schuster and Peter Wagner.
\newblock Mutual entrainment of two limit cycle oscillators with time delayed
  coupling.
\newblock {\em Prog. Theor. Phys.}, 81(5):939--945, 1989.

\bibitem{NewDDEBiftool}
J.~Sieber, K.~Engelborghs, T.~Luzyanina, G.~Samaey, and D.~Roose.
\newblock {\em {DDE-BIFTOOL} Manual --- Bifurcation analysis of delay
  differential equations}.
\newblock 2015.
\newblock {a}rXiv preprint arXiv:1406.7144; available at:
  \url{https://sourceforge.net/projects/ddebiftool}.

\bibitem{sonxu12}
Yongli Song and Jian Xu.
\newblock Inphase and antiphase synchronization in a delay-coupled system with
  applications to a delay-coupled {FitzHugh--Nagumo} system.
\newblock {\em IEEE Trans Neural Netw Learn Syst}, 23(10):1659--1670, 2012.

\bibitem{takfuj00}
Atsuko Takamatsu, Teruo Fujii, and Isao Endo.
\newblock Time delay effect in a living coupled oscillator system with the
  plasmodium of physarum polycephalum.
\newblock {\em Physical Review Letters}, 85(9):2026, 2000.

\bibitem{TerrienPRA17}
Soizic Terrien, Bernd Krauskopf, Neil G.~R. Broderick, Louis Andr{\'{e}}oli,
  Foued Selmi, R{\'{e}}my Braive, Gr{\'{e}}goire Beaudoin, Isabelle Sagnes, and
  Sylvain Barbay.
\newblock Asymmetric noise sensitivity of pulse trains in an excitable
  microlaser with delayed optical feedback.
\newblock {\em Physical Review A}, 96(4):043863, oct 2017.

\bibitem{Terrien2018OL}
Soizic Terrien, Bernd Krauskopf, Neil~GR Broderick, R{\'e}my Braive,
  Gr{\'e}goire Beaudoin, Isabelle Sagnes, and Sylvain Barbay.
\newblock Pulse train interaction and control in a microcavity laser with
  delayed optical feedback.
\newblock {\em Optics Letters}, 43(13):3013--3016, 2018.

\bibitem{wedslo21}
Kyle~CA Wedgwood, Piotr S{\l}owi{\'n}ski, James Manson, Krasimira
  Tsaneva-Atanasova, and Bernd Krauskopf.
\newblock Robust spike timing in an excitable cell with delayed feedback.
\newblock {\em Journal of the Royal Society Interface}, 18(177):20210029, 2021.

\bibitem{weiern14}
Lionel Weicker, Thomas Erneux, Lars Keuninckx, and Jan Danckaert.
\newblock Analytical and experimental study of two delay-coupled excitable
  units.
\newblock {\em Phys. Rev. E}, 89:012908, Jan 2014.

\bibitem{yanper09}
Serhiy Yanchuk and Przemyslaw Perlikowski.
\newblock Delay and periodicity.
\newblock {\em Phys Rev E}, 79(4):046221, 2009.

\bibitem{yeustr99}
MK~Stephen Yeung and Steven~H Strogatz.
\newblock Time delay in the {K}uramoto model of coupled oscillators.
\newblock {\em Phys Rev Lett}, 82(3):648, 1999.

\end{thebibliography}

\end{document}